\documentclass[fleqn,usenatbib]{mnras}

\usepackage{newtxtext,newtxmath}

\usepackage[T1]{fontenc}
\usepackage{ae,aecompl}

\usepackage{graphicx}	
\usepackage{amsmath}	
\usepackage{amssymb}	

\DeclareMathOperator*{\argmin}{argmin}
\title[Spin-ellipticity radial behaviour]{The SAMI Galaxy Survey: rules of behaviour for spin-ellipticity radial tracks in galaxies
}

\author[A. Rawlings et al.]{Alexander Rawlings,$^{1,2}$\thanks{E-mail: alexander.rawlings@helsinki.fi} Caroline Foster,$^{1,3}$ Jesse van de Sande,$^{1,3}$ Dan S. Taranu,$^{4,5}$ 
\newauthor Scott M. Croom,$^{1,3}$ Joss Bland-Hawthorn,$^{1,3}$  Sarah Brough,$^{6,3}$ Julia J. Bryant,$^{1,7,3}$
\newauthor Matthew Colless,$^{8,3}$ Claudia del P. Lagos,$^{5,3}$ Iraklis S. Konstantopoulos,$^{10}$ 
\newauthor Jon S. Lawrence,$^{9}$ \'Angel R. L\'opez-S\'anchez,$^{9,11}$ Nuria P. F. Lorente,$^{9}$ 
\newauthor Anne M. Medling,$^{12,8,13}$ Sree Oh,$^{8,3}$ Matt S. Owers,$^{11,14}$ Samuel N. Richards,$^{15}$ 
\newauthor Nicholas Scott,$^{1,3}$ Sarah M. Sweet$^{16,3}$ and Sukyoung K. Yi$^{17}$
\\
$^{1}$Sydney Institute for Astronomy, School of Physics, A28, The University of Sydney, NSW, 2006, Australia\\
$^{2}$Department of Physics, P.O. Box 64, FI-00014, University of Helsinki, Finland\\
$^{3}$ARC Centre of Excellence for All Sky Astrophysics in 3 Dimensions (ASTRO 3D)\\
$^{4}$Department of Astrophysical Sciences, Princeton University, 4 Ivy Lane, Princeton, NJ 08544, USA\\
$^{5}$International Centre for Radio Astronomy Research, 7 Fairway, The University of Western Australia, Crawley, Perth, WA 6009, Australia\\
$^{6}$School of Physics, University of New South Wales, NSW 2052, Australia\\
$^{7}$Australian Astronomical Optics, AAO-USydney, School of Physics, University of Sydney, NSW 2006, Australia\\
$^{8}$Research School of Astronomy and Astrophysics, Australian National University, Canberra, ACT 2611, Australia\\
$^{9}$Australian Astronomical Optics, AAO-MQ, Faculty of Science and Engineering, Macquarie University, NSW 2109, Australia\\
$^{10}$Atlassian 341 George St Sydney, NSW 2000\\
$^{11}$Department of Physics and Astronomy, Macquarie University, NSW 2109, Australia\\
$^{12}$Ritter Astrophysical Research Center University of Toledo Toledo, OH 43606, USA \\
$^{13}$Hubble Fellow\\
$^{14}$Astronomy, Astrophysics and Astrophotonics Research Centre, Macquarie University, Sydney, NSW 2109, Australia\\
$^{15}$SOFIA Science Center, USRA, NASA Ames Research Center, Building N232, M/S 232-12, P.O. Box 1, Moffett Field, \\ CA 94035-0001, USA\\
$^{16}$Centre for Astrophysics and Supercomputing, Swinburne University of Technology, PO Box 218, Hawthorn, VIC 3122\\
$^{17}$Department of Astronomy and Yonsei University Observatory, Yonsei University, Seoul 03722, Republic of Korea\\
}

\pubyear{2019}

\begin{document}
\label{firstpage}
\pagerange{\pageref{firstpage}--\pageref{lastpage}}
\maketitle

\begin{abstract}
We study the behaviour of the spin-ellipticity radial tracks for 507 galaxies from the Sydney AAO Multi-object Integral Field (SAMI) Galaxy Survey with stellar kinematics out to $\geq1.5R_\text{e}$. We advocate for a morpho-dynamical classification of galaxies, relying on spatially-resolved photometric and kinematic data. We find the use of spin-ellipticity radial tracks is valuable in identifying substructures within a galaxy, including embedded and counter-rotating discs, that are easily missed in unilateral studies of the photometry alone. Conversely, bars are rarely apparent in the stellar kinematics but are readily identified on images. Consequently, we distinguish the spin-ellipticity radial tracks of seven morpho-dynamical types: elliptical, lenticular, early spiral, late spiral, barred spiral, embedded disc, and 2-sigma galaxies. The importance of probing beyond the inner radii of galaxies is highlighted by the characteristics of galactic features in the spin-ellipticity radial tracks present at larger radii. The density of information presented through spin-ellipticity radial tracks emphasises a clear advantage to representing galaxies as a track, rather than a single point, in spin-ellipticity parameter space.
\end{abstract}




\section{Introduction}

 
 For the best part of the last century, galaxies have been studied solely using images. The classification scheme based on visual morphology of galaxy images first proposed by \citet{Hubble26} is still widely used. Elliptical (E) and lenticular (S0) galaxies are often referred to as early-types and barred or unbarred spirals (SB or S) known as late-types. Spirals may be further divided into early spirals (eSp) and late spirals (lSp). 
 
 In parallel, relatively recent advances in spatially-resolved spectroscopy, in particular integral field spectroscopy (IFS), has enabled the study of the stellar and gas kinematics of galaxies. Galaxy classification schemes have consequently been refined with both detailed light profile analyses \citep[e.g.][]{deVaucouleurs59}, and kinematics \citep[e.g.][]{Emsellem07, vandeSande17a, Cappellari11}. 
 

At first, it was assumed that early-type galaxies were pressure-supported. It was soon realised that E galaxies are not solely pressure-supported and hence often do rotate \citep[e.g.][]{BertolaCapaccioli75,ScorzaBender95}, with the majority of E galaxies exhibiting significant rotation \citep[e.g.][]{Emsellem11}, a trait initially thought to pertain only to later-types (i.e. S0 and S).

The Spectroscopic Aerial Unit for Reasearch on Optical Nebulae (SAURON) team \citep{Emsellem07} pioneered the use of a luminosity-weighted spin parameter, which is analogous to that commonly used in theoretical work, to classify early-type galaxies according to their kinematics. \citep{Emsellem07} proposed the spin parameter, represented as $\lambda_\text{R}$, conveniently condenses the 2D-kinematic information within a set aperture into a single parameter. When plotted as a function of mean apparent ellipticity, $\lambda_\text{R}$ is typically used to delineate early-type galaxies into two kinematic families: fast rotators and slow rotators. 

Kinematic classification has not supplanted morphological classifications however, in part because kinematic maps are observationally costly compared to images, making morphological classifications more readily available. Additionally, is there information to be gleaned from imaging that kinematic maps alone cannot provide? The field-of-view, spatial resolution and achievable depth of integral field spectroscopic data typically limit the radial extent at which a kinematic classification can be made, with typical kinematic classifications being based on the inner effective radius (i.e. $R\lesssim1R_\text{e}$) only \citep[e.g.][]{Emsellem07,Cappellari16a,vandeSande17a}. Hence, kinematic classifications are typically based on only about half the stellar light.

The implications of the limitations on the radial extent of IFS data became apparent as spatially-resolved stellar kinematics at large galactocentric radii became available. \citet{Coccato09} and \citet{Pulsoni17} identified a number of massive early-type galaxies with a marked decrease in rotational support beyond $1R_\text{e}$ using planetary nebulae as kinematic tracers. Similarly, using data from the SAGES Legacy Unifying Globulars and GalaxieS (SLUGGS) Survey \citep{Brodie14}, \citet{Proctor09} showed that although NGC~821 and NGC~2768 have comparable spins ($\lambda_\text{R}$, e.g. \citealt{Emsellem07}) at $1R_\text{e}$, their spins differ markedly in the outskirts. \citet{Weijmans09} subsequently confirmed the decreasing rotational support of NGC~821 at large radii using the SAURON spectrograph. \citet{Arnold11} found a significant fraction (6 out of 22) of early-type galaxies with decreasing outer rotational support. This feature of radially decreasing rotational support was interpreted as rotationally-supported ``embedded disc'' structures within larger scale slowly or non-rotating structures, whereby the disc dominates the flux within the inner radii (type ES, e.g. \citealt{Liller66}). Embedded discs have also been found in the SAMI Galaxy survey \citep{Foster18} in proportions consistent with those found in SLUGGS \citep{Arnold14,Bellstedt17a} once accounting for observational limitations.

\citet{SavorgnanGraham16} pointed out that morphological structures reminiscent of embedded disc galaxies have been known from photometry alone for decades \citep[e.g.][]{Liller66, Rix90, Cinzano94}, with some having been confirmed spectroscopically using stellar kinematics \citep[e.g.][]{Nieto88}. More recently, \citet{Jerjen00} found residuals with clear spiral structures embedded within the carefully subtracted bulge of a dwarf elliptical. 
Embedded disc galaxies are rarely discussed and often discarded, with many considering their structure (small scale disc in large scale bulge) ``unphysical'' \citep[e.g.][]{Allen06,Vika14}.
The question of how common embedded disc galaxies are needs to be settled in order to understand their importance in galactic evolution. 
Identification of embedded disc galaxies in datasets with limited radial coverage and spatial resolution is challenging \citep[e.g.][]{Foster18}.
\citet{Graham17} suggested that embedded disc galaxies should be readily identified in the spin-ellipticity diagram, a parameter space which combines both kinematic and photometric properties, through tell-tale counterclockwise radial tracks. This combined spectroscopy and imaging approach to identifying embedded disc galaxies has already been validated in \citet{Bellstedt17b}.

In this work, we further test the hypothesis that spin-ellipticity radial tracks can be used to identify embedded discs. We extend the method to all morphological types and look for ``rules of behaviour'' common to specific classes of galaxies\footnote{The title of this paper is borrowed from the seminal paper by Frank \citet{Briggs90}, who studied the rules of behaviour for HI warps using radial tracks of the orientation of the angular momentum vector in galaxies.}. We show that this combined approach of using both photometric and kinematic data provides a useful and harmonizing scheme to classify galaxies into meaningful (i.e. both visually and kinematically legitimate) categories. The paper is divided as follows: our data and sample selection are presented in Section \ref{sec:data}. Section \ref{sec:analysis} presents the analysis method used to extract and interpret the spin-ellipticity radial tracks for specific classes of galaxies and also outlines our results. A detailed discussion of these results can be found in Section \ref{sec:discussion}. A brief summary and our conclusions are outlined in Section \ref{sec:conclusions}.

We assume a $\Lambda$CDM cosmology with  $\Omega_{\rm m}=0.3$, $\Omega_{\lambda}$~$=$~$0.7$ and $H_0=70$ km s$^{-1}$ Mpc$^{-1}$.

\begin{figure*}
    \includegraphics[width=\textwidth]{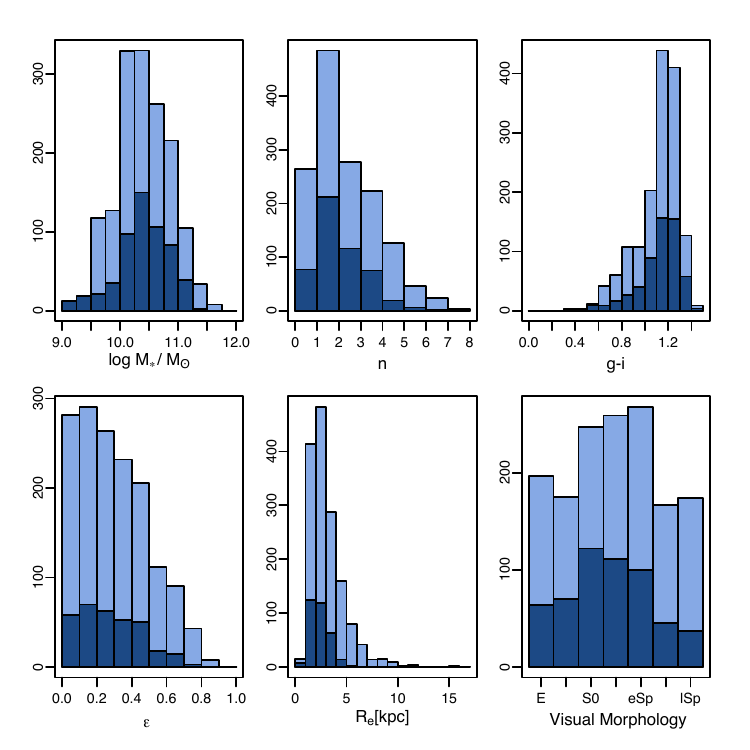}
    \caption{Histograms showing the distributions in stellar mass $\log M_*/M_\odot$, S\'ersic Index $n$, $(g-i)$ colour, apparent ellipticity $\varepsilon$, effective radius $R_\text{e}$, and visual morphology (E: elliptical, S0: lenticular, eSp: early spiral, and lSp: late spiral). The distributions are found from the SDSS imaging used by \citet{Cortese19} in identifying visual morphology. The parent sample is shown in light blue, with the study sample overlayed in dark blue. The study sample has no galaxies with $R_\text{e}>6$, a higher fraction of galaxies with high S\'ersic index, a higher fraction of galaxies with high $(g-i)$ colour and earlier morphological types than the parent sample.}
    \label{fig:samplehist}
\end{figure*}

\section{Data}\label{sec:data}
We use Sydney AAO Multi-object Integral Field \citep[SAMI;][]{Croom12} spectrograph data obtained as part of the SAMI Galaxy Survey \citep{Bryant15,Owers17}. The SAMI instrument boasts 13 integral field unit (IFU) hexabundles, enabling the aquisition of spatially-resolved spectroscopy for 13 objects simultaneously \citep[][]{BlandHawthorn11,Bryant14}. For the SAMI Galaxy Survey, we target 12 galaxies and one calibrator star per field. Each hexabundle is comprised of 61 tightly packed optical fibres offering a 73\% filling factor over a 15 arcsec diameter. In addition to the 13 hexabundles, the SAMI instrument has 26 individual sky fibres over its one degree field-of-view. Optical fibres (hexabundle and sky) are fed into the AAOmega spectrograph \citep{Sharp06} on the 3.9m Anglo-Australian Telescope. 

The final SAMI Galaxy Survey sample contains $\sim3000$ low redshift ($0.004\le z\le0.095$) galaxies with a step-wise redshift-dependent stellar mass selection. The details of the SAMI target selection for the Galaxy And Mass Assembly (GAMA, \citet{Driver11}) fields and the cluster samples are discussed by \citet{Bryant15} and \citet{Owers17}, respectively. The early, first and second public data releases are described in \citet{Allen15}, \citet{Green18} and \citet{Scott18}, respectively.

The 580V and 1000R gratings are used for the blue and red arms of the AAOmega spectrograph. This yields a median spectral resolution of $R\sim1809$ and $R\sim4310$ in the blue and red arms, respectively, for a broad combined wavelength coverage of 3750-5750 \AA\ (blue) and 6300-7400 \AA\ (red). The full SAMI data reduction process is described in \citet{Allen15} and \citet{Sharp15}, and summarised here. We use the {\sc 2dFDR} pipeline \citep{Croom04} to reduce the SAMI data. The pipeline performs the common spectral reduction steps: subtraction of bias frames, flat fielding, cosmic ray removal, wavelength calibration using CuAr arc frames and sky subtraction. The technique developed by \citet{Sharp10} is used to extract a spectrum for each fibre. Details of the spectral data reduction steps performed by {\sc 2dFdr} can be found in \citet{Hopkins13}. Flux calibration is then performed using a primary standard star observed on the same night as the observations. A secondary standard star observed simultaneously with each field is used for flux scaling and telluric absorption corrections. The observing strategy involves a seven point dither pattern to deal with gaps between fibres within the hexabundles and ensure continuous sampling. A minimum of 6 dithers is required to reconstruct datacubes for any galaxy. Flux and covariance are carefully propagated onto a grid as described in \citet{Sharp15}.

\subsection{Extraction of the spatially-resolved kinematics}
For each spaxel, the line-of-sight velocity distribution (LOSVD) is parametrised with a Gaussian using the penalised pixel-fitting ({\sc pPXF}, \citealt{Cappellari04, Cappellari16b}) algorithm. The recession velocity, $V$, and velocity dispersion, $\sigma$, are measured by fitting template spectra. As described in \citet{vandeSande17a}, spectra within elliptical annuli are first combined into a high signal-to-noise spectrum to determine the best template to be used for fitting the LOSVD on individual member spaxels. This two-step fitting process is used to mitigate uncertainties due to template mismatch, which significantly affect lower signal-to-noise spectra. For each spaxel, residuals between the observed spectrum and the pre-determined template broadened through convolution with a Gaussian LOSVD ($V$, $\sigma$) are minimised.

\subsection{Visual classification and image preparation}
Visual classification of galaxy morphology of the entire SAMI Galaxy Survey is conducted on SDSS and VST images, as described in \citet{Cortese19}. Image cutouts for all SAMI targets are made from either the Kilo Degree Survey \citep[KiDS;][]{deJong15,deJong17} observed with the European Southern Observatory's Very Large Telescope (VLT) Survey Telescope OmegaCAM, or the Subaru Telescope Hyper Suprime Camera (HSC) Public Data Release 1 \citep[PDR1;][]{Aihara18}.
The KiDS $r$-band co-added images are downloaded directly from the European Southern Observatory archive. The images are $5\times360$s dithers for a total exposure time of 1800s. The KiDS cutouts are 300$\times$300 pixels with a pixel scale of 0.2 arcsec/pix.
HSC PDR1 $r$-band images are downloaded using the PDR1 \citep{Aihara18} ``DAS Quarry"
tool\footnote{\url{https://hsc-release.mtk.nao.ac.jp/das_quarry/}}. All HSC images are taken from the Wide Survey with 600s total exposure time.
Both surveys report typical seeing around 0.68 arcsec. Cutouts are nominally 1 arcmin $\times$ 1 arcmin centered on the SAMI targets positions. In practice, the HSC pixel scale is 0.168 arcsec/pix, hence cutouts of $360\times360$ pixels yield images that are 1.008 arcmin on the side. 


\subsection{Sample selection}
Our sample selection follows \citet{vandeSande17a} and \citet{Foster18}, but only includes the GAMA sample of the SAMI Galaxy Survey. We do not include the cluster sample because the imaging quality (i.e. depth and spatial resolution) is not suitable for our analysis. We select from GAMA galaxies included in the internal team release version v0.11. First, galaxies with no spaxels satisfying $V_{\rm error}<30$ km s$^{-1}$ and $\sigma_{\rm error} < 0.1\sigma + 25$ km s$^{-1}$ are discarded. We further require that the spatial resolution or half-width-at-half-maximum (HWHM) of the SAMI data not exceed 1$R_\text{e}$. From the remaining parent sample of 2102 galaxies with reliable stellar kinematics, we identify a sample of galaxies with measured stellar kinematics ($V$ and $\sigma$) out to at least 1.5$R_\text{e}$. The outermost radius probed must have at least 85 percent of spaxels meeting our $V_{\rm error}$ and $\sigma_{\rm error}$ quality criteria. There are 568 SAMI galaxies satisfying this selection. During the analysis of the sample, a total of 51 galaxies fail to meet the analysis requirements imposed (Section \ref{sec:analysis}), and are thus removed from the sample. The removal of the final 51 galaxies reduces the total number of galaxies studied to 507. As discussed in \citet{Foster18}, this selection is biased in favour of earlier morphological types and higher S\'ersic index values, demonstrated in Fig. \ref{fig:samplehist}. By construction, our selection removes galaxies with effective radii larger than two thirds the size of an individual hexabundle. 


\begin{figure*}
	\includegraphics[width=\textwidth]{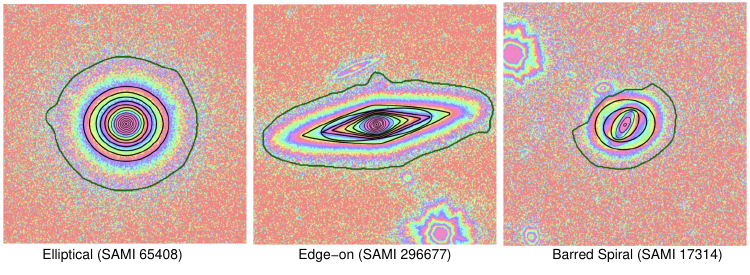}
    \caption{Ellipse fitting to an elliptical galaxy (first panel), an edge-on galaxy (second panel), and to a barred spiral galaxy (third panel). The isophotal lines for the inner 90\% of ellipses (solid black lines) are superimposed on either the KiDS or HSC image of the galaxy using the {\sc ProFound} package developed by \citet{Robotham18}. The outer 10\% of ellipses are not used in the analysis due to low S/N ratios. The area segmented from the original image corresponding to the study galaxy is enclosed by the green border. Note the highly-elliptical ellipses at small radii in the isophotal fit to the barred spiral galaxy SAMI 17314, and the return to less-elliptical ellipses at outer radii.}
    \label{fig:ellipsefit}
\end{figure*}

\begin{figure*}
	\includegraphics[width=\textwidth]{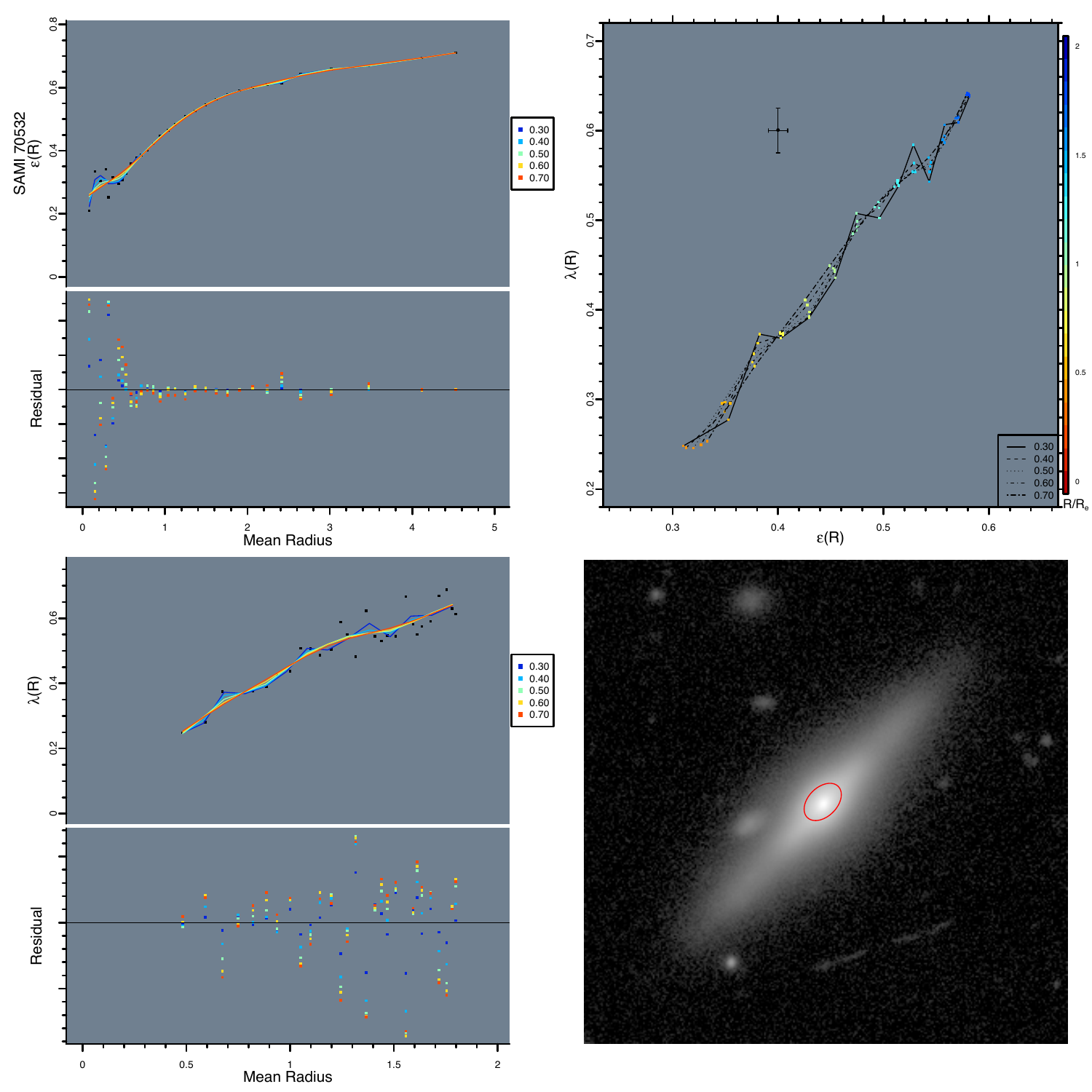}
    \caption{Differing choices of the {\sc R} spline smoothing parameter (corresponding to differing minimised $\kappa$ in Eq. \ref{eq:spline}) for the spline fits for ellipticity (top left), spin (bottom left), and the resulting spin-ellipticity plot (top right) for SAMI 70532. The KiDS image is presented (bottom right). The residuals for ellipticity and spin are given below the respective plot. Increasing the smoothing spline parameter increases the smoothing effect, potentially losing important data (red spline), whereas decreasing the smoothing parameter increases the likelihood of fitting to noise (dark blue spline). A midway smoothing parameter value of 0.5 is chosen as a compromise between over-fitting and over-smoothing for later analysis. As demonstrated in the spin-ellipticity radial plot, the resulting effect on the spin-ellipticity radial track is qualitatively minimal. The error for both $\varepsilon(R)$ and $\lambda(R)$ corresponding to a smoothing parameter value of 0.5 is given in the spin-ellipticity radial plot. }
    \label{fig:splinecomp1}
\end{figure*}

\begin{figure*}
	\includegraphics[width=\textwidth]{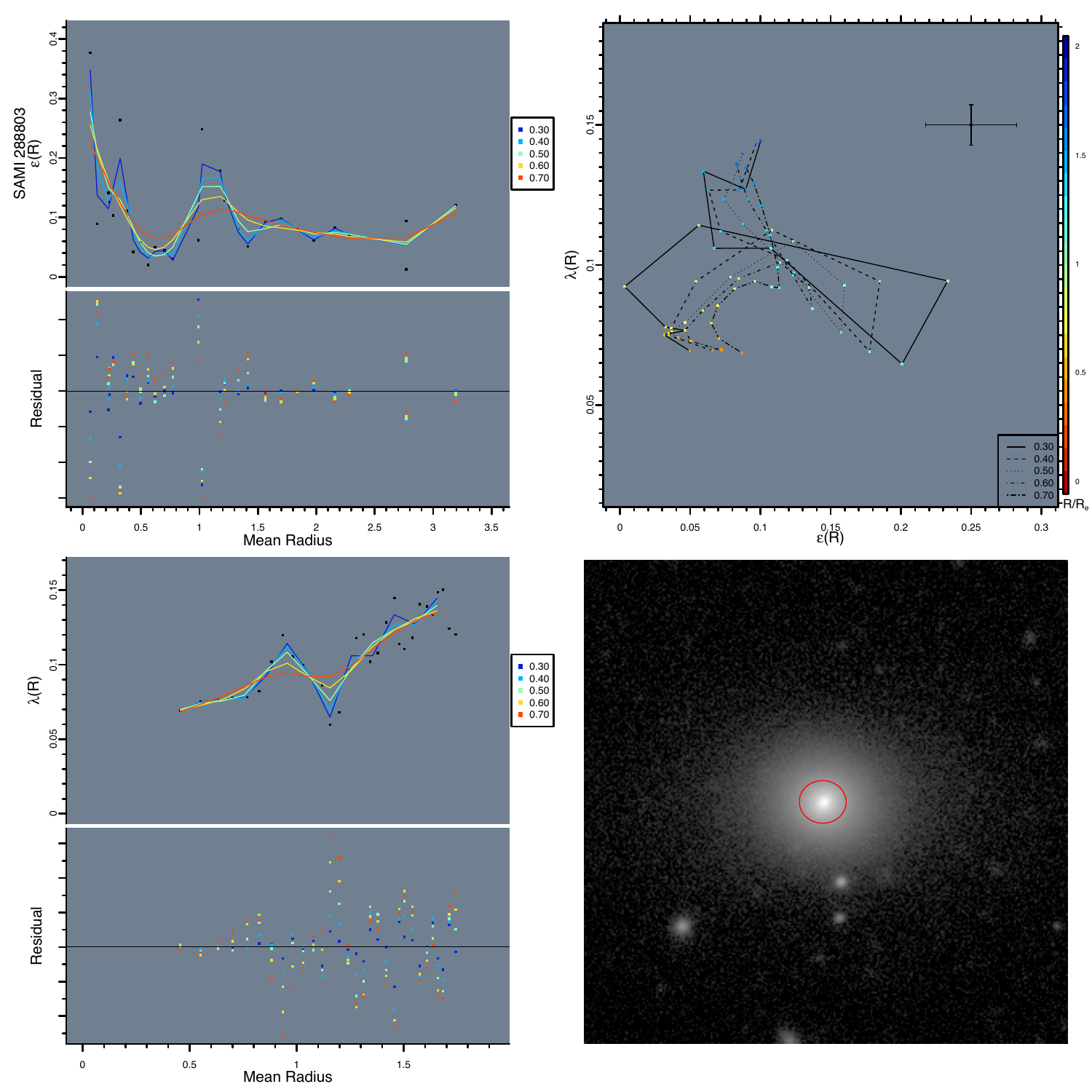}
    \caption{Subplots as per Fig. \ref{fig:splinecomp1} for SAMI 288803. SAMI 288803 demonstrates a minority case whereby the choice of the spline smoothing parameter (and resulting $\kappa$ in Eq. \ref{eq:spline}) can drastically affect the spin-ellipticity radial track of the galaxy. A smoothing parameter value of 0.3 fits primarily to noise in the data, whereas a smoothing parameter value of 0.7 smooths over data that potentially has physical meaning. As was done for SAMI 70532, a compromise between over-fitting and over-smoothing was made by selecting a smoothing parameter value of 0.5. Note the differing scale for $\varepsilon(R)$ and $\lambda(R)$ for SAMI 288803 compared to SAMI 70532.}
    \label{fig:splinecomp2}
\end{figure*}

\begin{figure*}
	\includegraphics[width=\textwidth]{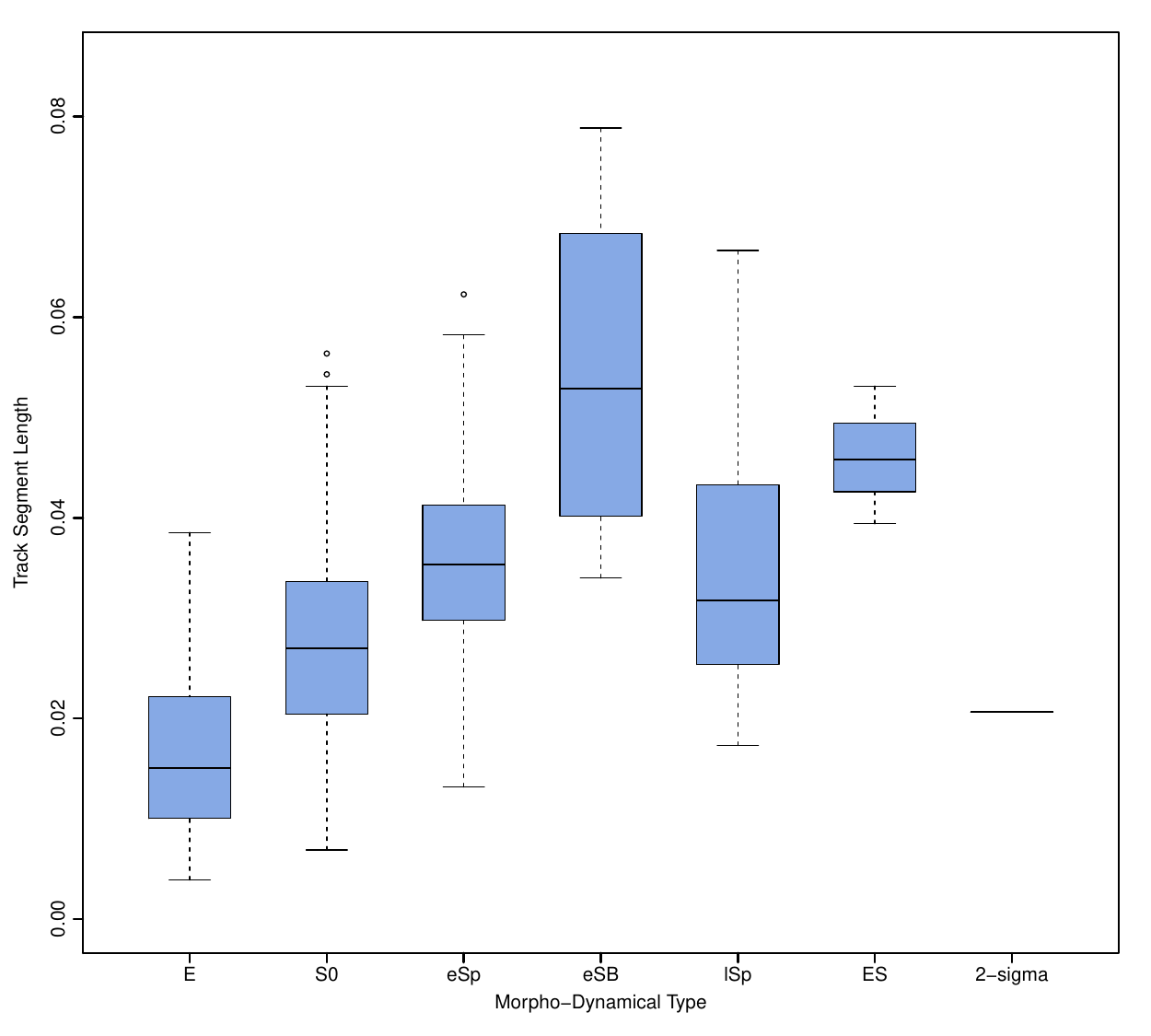}
    \caption{Box plots of track segment length, defined as the mean segment length between adjacent radial points in spin-ellipticity parameter space, for each morpho-dynamical type. The whiskers extend to include those data points within 1.5$IQR$ above and below the box, and outliers beyond 1.5$IQR$ are represented by individual circles. In spin-ellipticity parameter space, galaxies with a short track segment length have closely-spaced radial points, and galaxies with long track segment lengths have far-spaced radial points. Elliptical galaxies have the smallest median track segment length, and barred spirals the largest. For the 2-sigma morpho-dynamical type, only one data point exists (SAMI 239560). }
    \label{fig:tracklength}
\end{figure*}

\begin{figure*}
    \includegraphics[width=\textwidth]{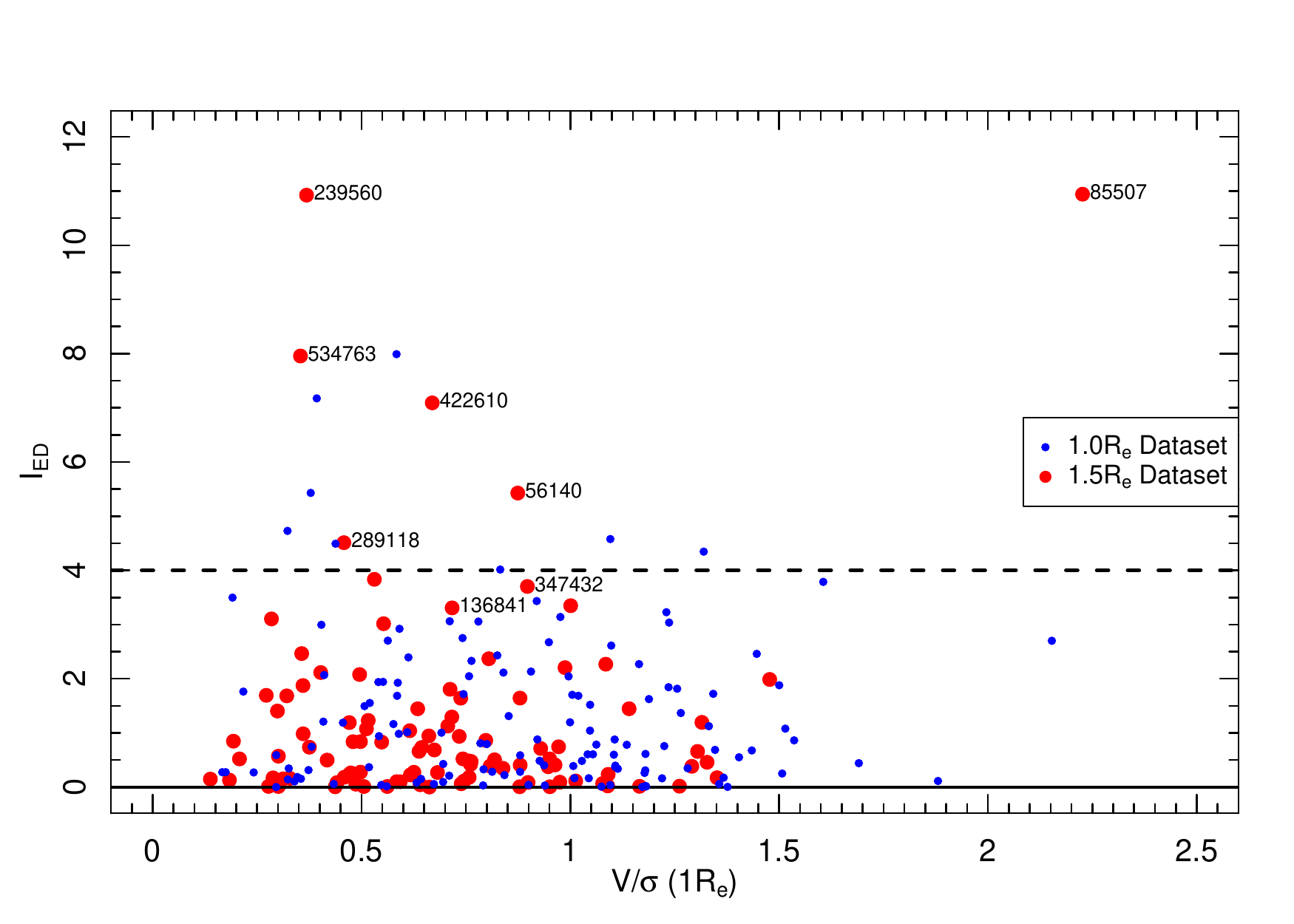}
    \caption{The embedded disc index, $I_\text{ED}$, plotted as a function of $V/\sigma$ at 1$R_\text{e}$ for those galaxies with $I_\text{ED}>0$. Galaxies with kinematic data out to 1.5$R_\text{e}$ are plotted in red, and galaxies with kinematic data out to only 1.0$R_\text{e}$ in blue. The dashed horizontal line at $I_\text{ED}=4$ indicates the threshold above which visual identification of embedded discs may be made, as per \citet{Foster18}. Galaxies of the $1.5R_\text{e}$ dataset with $I_\text{ED}>4$ are labelled, as are the embedded disc galaxies SAMI 56140, SAMI 347432, and SAMI 136841.}
    \label{fig:EDI_plot}
\end{figure*}

\begin{figure*}
	\includegraphics[width=\textwidth]{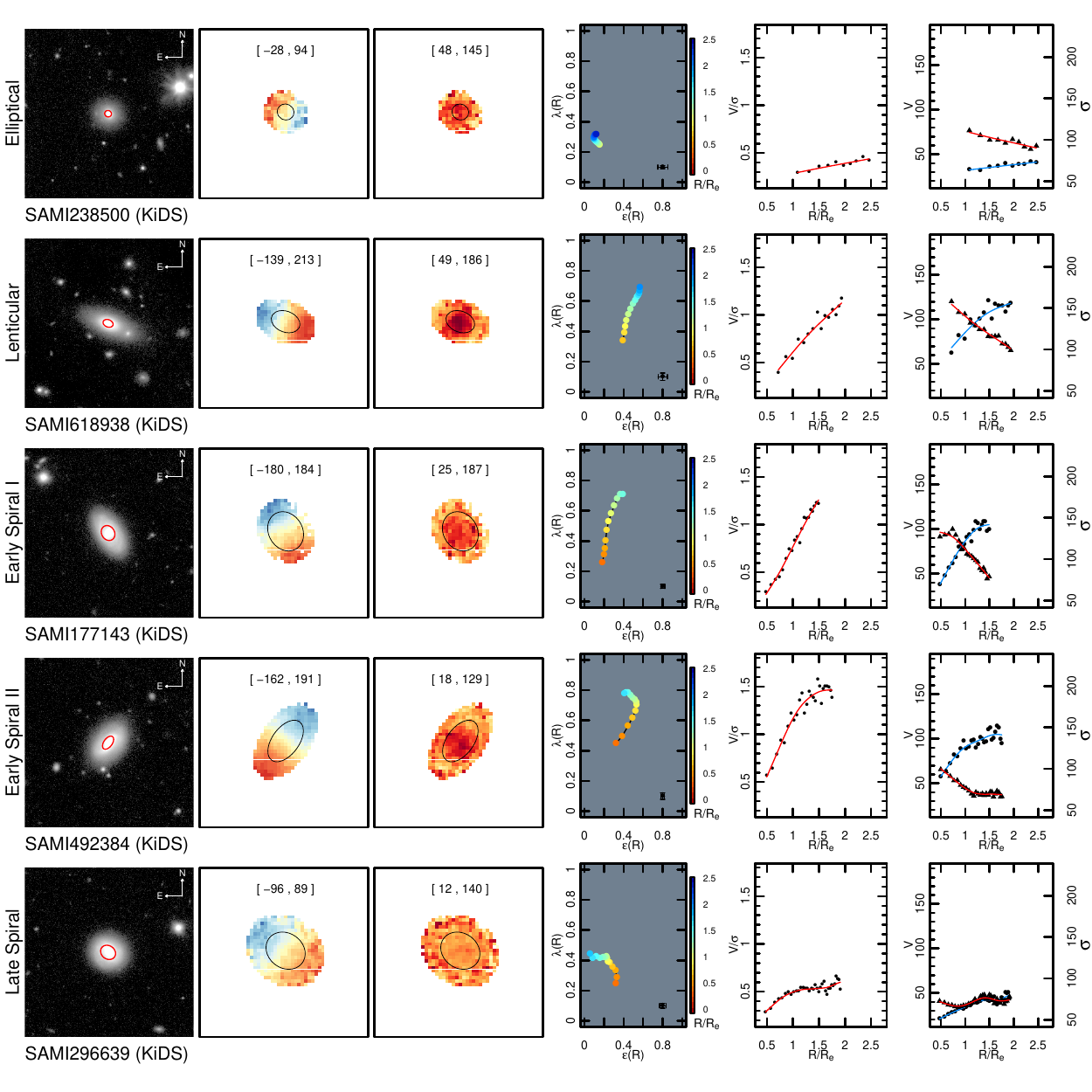}
    \caption{The KiDS $r$-band image for an example of each morpho-dynamical type with the stellar recession velocity ($V$ [km/s], second column) and stellar velocity dispersion ($\sigma$ [km/s], third panel) maps. The red ellipse (column one) and the black ellipse (columns two and three) indicate 1$R_\text{e}$. Recession velocity and velocity dispersion ranges are given in the format [blue, red] and [yellow, red] in the respective panels. In the fourth column are the spin-ellipticity radial tracks with increasing radius from $R_\text{min}$ to $R_\text{max}$ in 0.1$R_\text{e}$ increments. The points are coloured red to blue, with the colour of the point indicating the radius in $R_\text{e}$ probed. Typical scatter in $\varepsilon(R)$ and $\lambda(R)$ are shown as error bars on the bottom right. Also shown are the $V/\sigma$ profiles with the spline fit as a red solid line (fifth column) and combined $V$ (left axis labels, blue line spline and circle symbols) and $\sigma$ (right axis labels, red line spline and triangle symbols) profiles (right-most panel). Example elliptical (E, top row), lenticular (S0, second row), early spirals (eSp, third and fourth rows) and late spiral (lSp, bottom row) are shown. Different morpho-dynamical types show a range of typical spin-ellipticity radial tracks.}
    \label{fig:morphology_gals}
\end{figure*}

\begin{figure*}
	\includegraphics[width=\textwidth]{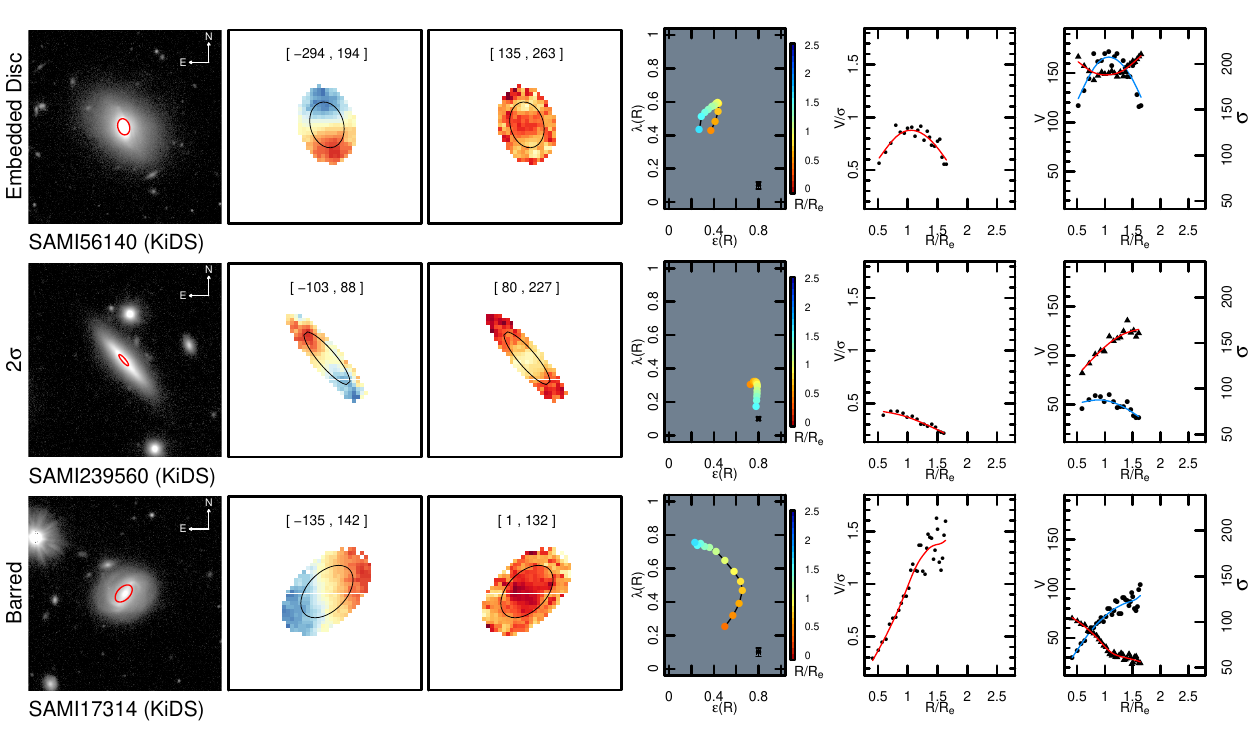}
    \caption{Columns as per Fig. \ref{fig:morphology_gals} for a typical embedded disc (ES, top row), $2\sigma$ (second row), and early barred spiral (eSB, third row) in SAMI. These special classes of galaxies show tell-tale patterns in their spin-ellipticity tracks.}
    \label{fig:special_gals}
\end{figure*}

\begin{figure*}
	\includegraphics[width=\textwidth]{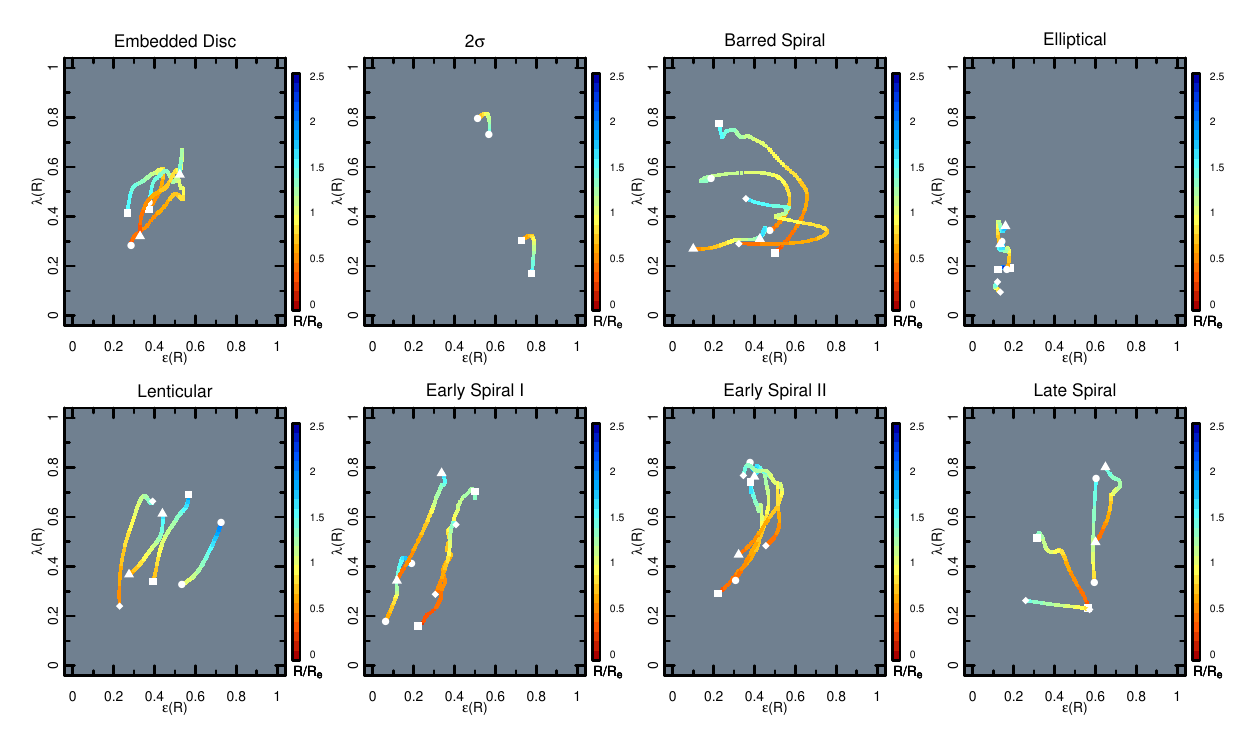}
    \caption{Spin-ellipticity radial tracks for a maximum of four galaxies of each morpho-dynamical type with stellar kinematics measured out to 1.5$R_\text{e}$. The radial range probed is indicated by the point colour on the plot, with a normalised radius of 0 indicated by dark blue, and the maximum 2.5 by deep red (this is the same scaling as in Figs. \ref{fig:morphology_gals} and \ref{fig:special_gals}). The typical radial range probed extends to $\sim$1.5$R_\text{e}$. Note early spirals are divided across two panels, to better highlight the difference between Type I and Type II early spiral galaxies. Whilst four galaxies of each type are shown for visual clarity, the spin-ellipticity radial tracks of all galaxies in our sample may be found in the supplementary material.}
    \label{fig:tracks}
\end{figure*}

\section{Analysis and results}\label{sec:analysis}

In this section we present our analysis and findings from the sample of 507 SAMI galaxies satisfying our selection criteria of those galaxies with measured stellar kinematics out to $\gtrsim1.5R_\text{e}$. 

\subsection{Kinematic profiles}

We apply a clipping to the velocity and velocity dispersion maps to remove kinematic outlying spaxels that are beyond three standard deviations from the mean spaxel value. For those galaxies that had a spaxel clipped, the kinematic profiles are recalculated. Next, we determine radial variation in the stellar kinematics of the selected SAMI galaxies by computing the local spin parameter $\lambda(R)$, originally defined in \citet{Emsellem07} and adapted by \citet{Bellstedt17b}:
    \begin{equation}\label{eq:spinloc}
        \lambda(R)=\frac{\sum_{i\in A}R_i|V_i|}{\sum_{i\in A}R_i\sqrt{V_i^2+\sigma_i^2}},
    \end{equation}
where $R_i$ is the galactocentric radius, $V_i$ the normalised recession velocity, and $\sigma_i$ the velocity dispersion, within the $i^\text{th}$ spaxel in the elliptical annulus $A$ with circularised radius $R$. In contrast with previous work on spin-ellipticity radial tracks, we measure $R$ using a single global ellipticity value. This is necessary as the kinematic maps have significantly lower spatial resolution (FWHM $\sim2$ arcsec) and sampling (0.5 arcsec pixel scale) compared to the HSC (0.168 arcsec pixel scale) and KiDS (0.2 arcsec pixel scale) imaging sampling and resolution ($\sim0.6$ arcsec). Allowing a radially varying ellipticity value when measuring $\lambda(R)$ could lead to the same kinematic pixel being included in non-consecutive radial bins. Hence, adopting a single global ellipticity value, i.e. sampling along concentric ellipses, ensures individual pixels are only included in consecutive rolling bins.

We calculate the ratio of ordered to random stellar motion as a function of radius following e.g. \citet{Wu14} and \citet{Foster18}:
    \begin{equation}\label{eq:vsigloc}
        V/\sigma(R)=\sqrt{\sum_{i\in A}\left(\frac{V_i}{\sigma_i}\right)^2}.
    \end{equation}
Rolling annular bins of 25 spaxels at ten spaxel intervals are chosen for calculating both $\lambda(R)$ and $V/\sigma(R)$. We find that $\lambda(R)$ scales with $V/\sigma(R)$ as $\lambda(R)=\zeta\;v/\sigma(R)/\sqrt{1+\zeta^2\;v/\sigma(R)^2}$ as described by \citet{Emsellem11}. For a selection of radial measurements of $v/\sigma(R)$ and $\lambda(R)$ between 0.5$R_\text{e}$ and 2.0$R_\text{e}$, $\zeta$ is found to be $\simeq0.85$, similar to the value of $\zeta\simeq0.9$ reported in \citet{Bellstedt17b} for the SLUGGS survey.

\subsection{Photometry profiles}\label{ssec:photoprofile}
We fit ellipses extending from a common flux-weighted galactic centre along isophotes using the {\sc ProFound} \citep{Robotham18} package. The package {\sc ProFound} uses the watershed approach to segment an image, as described in \citet{Robotham18}. During the analysis, 43 galaxies within the sample cannot be correctly segmented due to low S/N and neighbouring structures adversely affecting image segmentation, and are removed from the study. The {\sc ProFound} package calculates the modified radius $R_m$, which for elliptical isophotes is:
    \begin{equation}\label{eq:rm}
        R_m=\sqrt{\left(R_t\sin{(\theta_t)A_\text{rat}}\right)^2+\left(R_t\cos{(\theta_t)}\right)^2},
    \end{equation}
with $A_\text{rat}=1-\varepsilon$ the minor-to-major axis ratio of each isophotal ellipse ($0\leq A_\text{rat} \leq 1$),
    \begin{equation}
        R_t=\sqrt{\left(x-x_\text{cen}\right)^2+\left(y-y_\text{cen}\right)^2}
    \end{equation}
and
    \begin{equation}
        \theta_t=\arctan{\frac{x-x_\text{cen}}{y-y_\text{cen}}+\theta}.
    \end{equation}
Hence, $x$ and $y$ specify the location of a pixel, centred at $x_\text{cen}$ and $y_\text{cen}$, and $\theta$ is the major-axis position angle \citep{Robotham17}. Some galaxies are better fitted with boxy isophotes. Consequently, $R_m$ is adjusted to:
\begin{equation}\label{eq:rmbox}
    R_m=\left[\left(R_t\sin(\theta_t)A_\text{rat}\right)^{(2+B)}+\left(R_t\cos(\theta_t)\right)^{(2+B)}\right]^{\frac{1}{2+B}}
\end{equation}
where $B$ is the boxiness, defined to be $-1\leq B \leq1$ in the \citet{Robotham17} software package. If the galaxy is not described by a boxy profile, $B$ is 0, and thus Eq. \ref{eq:rmbox} reduces to Eq. \ref{eq:rm}. Both a boxy and non-boxy fit are created for each galaxy image, and the better fit is determined through visual inspection. For the 188 galaxies with imaging from both HSC and KiDS, we select the fit with the higher number of elliptical contours (typically the HSC image), provided that the fit is correctly segmented. If the fit with the higher number of elliptical contours is incorrectly segmented (as a result of e.g. bright nearby object), the fit with the lower number of elliptical contours (typically KiDS) is selected, provided that the fit is correctly segmented. In the event that neither the HSC nor the KiDS image are correctly segmented, the galaxy is removed from the sample. 

\subsection{Spin-ellipticity radial tracks}\label{ssec:sertracks}
A smoothing cubic spline is fit to the ellipticity as a function of radius for the inner 90\% of ellipses (indicated as solid black ellipses in Fig. \ref{fig:ellipsefit}), as the outer 10\% of ellipses are typically faint and noisy. A smoothing cubic spline is also fit to the local spin as a function of radius. The spline fit for both variables requires there to be a minimum of five radial points. A total of 18 galaxies fail to meet this criterion, and are discarded from further analysis. The smoothing cubic spline of a set of variables $\{x_i, y_i\}$ over some interval $[a, b]$ is defined in \citet{Green94} as:
\begin{equation}\label{eq:spline}
    \hat{f}=\argmin_{f}\sum^n_{i=1}\left(y_i-f(x_i)\right)^2+\kappa\int^b_a\left(f''(x)\right)^2dx.
\end{equation}
The criterion is a balance between least squares optimisation and a penalty term that increases as the second derivative of the fitted function $f$ introduces higher order moments of $x$. The $\kappa\geq0$ parameter determines the strength of each penalty term in the minimisation \citep{Green94}. We test the spline fits for a set of five splines per galaxy with differing $\kappa$ with the {\sc R} function {\sc smooth.spline} which takes as argument the desired smoothing parameter, from which the minimised penalty term $\kappa$ in Eq. \ref{eq:spline} is calculated. A lower smoothing parameter value allows the spline to more closely follow the data, whereas a larger smoothing parameter value will not follow the data as closely. Consequently, the choice of smoothing parameter is a balance between over-fitting the data (over-interpreting the noise) and potentially smoothing over physically meaningful features. 

We consider smoothing parameter values between 0.3 and 0.7 inclusive, in 0.1 increments. Creating five splines for both ellipticity and spin enables us to determine the most appropriate smoothing spline, and to test the robustness of the model against differing penalty values. A plot of the five splines used to model ellipticity and spin, and the resulting spin-ellipticity plot, is shown in Figs. \ref{fig:splinecomp1} \& \ref{fig:splinecomp2}. We find the qualitative trends in spin-ellipticity parameter space for each morpho-dynamical type are largely independent of the penalty value $\kappa$ chosen (e.g. Fig. \ref{fig:splinecomp1}), however a minority of spin-ellipticity radial tracks are sensitive to the chosen smoothing parameter (e.g. Fig. \ref{fig:splinecomp2}). For consistency, and to minimise both the effect of over-fitting and the effect of over-smoothing, a smoothing parameter value of 0.5 is chosen to construct the spin-ellipticity radial tracks for galaxies in the sample. 

 Spin-ellipticity radial tracks are limited to the radial extent that is common to both the spin and ellipticity profiles ($R_\text{min}$ to $R_\text{max}$). We thus define the term `track segment length', denoted $\overline{L}$ as being the mean length of the spin-ellipticity radial track of a galaxy between adjacent radial points in spin-ellipticity parameter space. A short track segment length will have closely-spaced radial points, and a long track segment length will have well-spread radial points on the spin-ellipticity radial plot. The track segment length is derived from the arc length in calculus:
\begin{equation}\label{eq:lineintegral}
    L=\int_{C}{\sqrt{(dx)^2+(dy)^2}ds}
\end{equation}
which, given the parametric equations used to construct the spin-ellipticity parameter space $\varepsilon(R)$ and $\lambda(R)$, becomes:
\begin{equation}\label{eq:tslF}
    L=\int_{C}{\sqrt{(d(\varepsilon(R)))^2+(d(\lambda(R)))^2}dR}.
\end{equation}
For a finite number $N$ of radial points $R_i$ spanning from $R_\text{min}$ to $R_\text{max}$ that form the set $\{R_i\in\mathbb{R}|i\in\mathbb{N}\land 1\leq i\leq N \land R_{i+1}-R_i=k\}$, the track segment length is thus:
\begin{equation}\label{eq:tsl}
    \overline{L}=\frac{\sum_{i=1}^{N-1}{\sqrt{(\varepsilon(R_{i+1})-\varepsilon(R_{i}))^2+(\lambda(R_{i+1})-\lambda(R_i))^2}}}{N-1}.
\end{equation}
The track segment length represents the average rate of change in both local spin and ellipticity with radius. We find no statistically significant correlation between track segment length and the number of radial points sampled.

The median radial range $R_\text{max}-R_\text{min}$ probed is similar across all morpho-dynamical types in the sample, ranging from 1.12$R_\text{e}$ for lSp galaxies to 1.29$R_\text{e}$ for S0 galaxies.

We find that ellipticals have the shortest median track segment length of the morpho-dynamical types, and barred spirals the longest median track segment length, as shown in Fig. \ref{fig:tracklength}. The morpho-dynamical types with the largest median track segment length have a pronounced bulge-disc transition: these are eSB, eSp, and ES galaxies. The use of the track segment length is a quantitative measure of the radial spacing of galaxy's spin-ellipticity radial track, directly related to its qualitative appearance in spin-ellipticity parameter space. 

We also compute the embedded disc index $I_\text{ED}$, as defined in \citet{Foster18}, for each galaxy:  
    \begin{equation}\label{eq:ied}
        I_\text{ED}=\frac{(V/\sigma)_\text{max}-V/\sigma(R_{\rm max})}{(V/\sigma)_\text{sd}}
    \end{equation}
with $(V/\sigma)_\text{max}$ the maximum and $V/\sigma(R_\text{max})$ the outermost value of the ratio of ordered to pressure dynamical support. The standard deviation about the $V/\sigma$ profile is $(V/\sigma)_\text{sd}$.
The embedded disc index is a measure of the likelihood of a galaxy possessing an embedded disc structure. As proposed by \citet{Foster18}, an $I_\text{ED}$ of $\simeq4$ corresponds to the threshold for visual identification of embedded discs, indicating a statistically significant decline in the profile (shown as a dashed horizontal line in Fig. \ref{fig:EDI_plot}). 

 We use the Hubble classification scheme as a prior, and extend the classes to include those galaxies with interesting kinematic features: namely embedded disc and counter-rotating disc galaxies (also known as 2-sigma). We find the kinematics of a galaxy are linked to its visual morphology. In Figs. \ref{fig:morphology_gals} and \ref{fig:special_gals}, we show a selection of typical diagnostic plots for each morpho-dynamical type with measured kinematic data out to 1.5$R_\text{e}$. Plots for those galaxies discussed in the text, but not included in the main figures, may be found in Appendix \ref{sec:app1p5}, while plots for all galaxies may be found in the supplementary material. 
 
 Elliptical galaxies demonstrate spin-ellipticity radial tracks in a compact section of spin-ellipticity space: $\varepsilon(R)\lesssim0.3$, $\lambda(R)\lesssim0.4$ in $>85\%$ of elliptical galaxies, even though the radial range probed extends to $\gtrsim1.9R_\text{e}$ in most ellipticals. Typically, elliptical galaxies have a short track segment length with a median value of $\overline{L}_\text{E}=0.015$ (Fig. \ref{fig:tracklength}). However, elliptical galaxies with significant rotation have a radially-increasing ellipticity and spin, thus tracing a slightly longer track segment length of up to $\overline{L}_\text{E}\simeq0.04$. Comparatively, elliptical galaxies with little or no significant rotation have constant ellipticity and little radial dependence on spin, resulting in spin-ellipticity radial tracks occupying low spin and low ellipticity parameter space (e.g. SAMI 238500 in Fig. \ref{fig:morphology_gals}). Elliptical galaxies with defined rotation display a radial decrease in velocity dispersion (Fig. \ref{fig:morphology_gals}). 
 In more than 90\% of elliptical galaxies, the $V/\sigma(R)$ profile monotonically increases with radius. 
 
 Conversely, lenticular, early spiral, and late spiral galaxies qualitatively display extended spin-ellipticity radial tracks, with the tracks of lenticular galaxies indicating an increase in spin and ellipticity with radius. The median track segment lengths of S0 and lSp galaxies are found to be $\overline{L}_\text{S0}=0.027$ and $\overline{L}_\text{lSp}=0.028$ respectively (Fig. \ref{fig:tracklength}). Early spirals tend to increased track segment lengths, with a median of $\overline{L}_\text{eSp}=0.038$. Approximately half of early spirals display a similar trend in their spin-ellipticity radial tracks to lenticulars, whilst other early spirals display a sudden radial decrease in ellipticity at outer radii. To further study the spin-ellipticity radial tracks of early spirals, we thus divide the early spiral population into two types: Type I, which display spin-ellipticity radial tracks qualitatively similar to lenticulars (Type I in Fig. \ref{fig:tracks}), and Type II, which display a sudden radial decrease in ellipticity at outer radii (Type II in Fig. \ref{fig:tracks}). The overall-linear slope between spin and ellipticity with radius typically has less fluctuation for lenticular galaxies than that of Type I early spirals. Late spirals do not demonstrate as clearly defined a trend as lenticular or early spirals, with the spin-ellipticity radial profile spanning a range of slopes, as shown in Fig. \ref{fig:tracks}. Clearly defined rotation is seen in all but four spiral galaxies (SAMI 55245, SAMI 296742, SAMI 422289, and SAMI 543895, further discussed in Section \ref{ssec:spiral}), and a radial decrease in velocity dispersion is displayed by the majority of early-type spirals, and the minority of late-type spirals (i.e. SAMI 177143 and SAMI 296639 respectively in Fig. \ref{fig:morphology_gals}). The $V/\sigma(R)$ profile increases with radius for spiral galaxies of both early- and late-types.  

We confirm that embedded disc galaxies, defined as galaxies with flux from the inner radii dominated by a disc component, and flux at the outer radii dominated by the bulge component \citep[i.e.][]{Foster18}, produce a distinct counter-clockwise signature in their spin-ellipticity radial tracks (shown in the top row of Fig. \ref{fig:special_gals}). The spin-ellipticity radial tracks of embedded disc galaxies in our sample are reminescent of those seen in the SLUGGS survey, discussed in \citet{Bellstedt17b}. From this trend, three embedded disc galaxies  (SAMI 56140 in Fig. \ref{fig:special_gals}, and SAMI 136841 and SAMI 347432) are clearly identified, and a further four potential embedded disc galaxies (SAMI 91697, SAMI 505341, SAMI 545903, and SAMI 618976) discovered; diagnostic plots can be found in Appendix \ref{sec:app1p5}. The decrease in ellipticity and spin occurs at differing radii for each embedded disc galaxy (Fig. \ref{fig:tracks}). The median track segment length of embedded disc galaxies is $\overline{L}_\text{ES}=0.034$ (Fig. \ref{fig:tracklength}). Embedded disc galaxies display an initial increase in $V/\sigma(R)$, peaking at $\sim1$-$1.5R_\text{e}$, followed by a decrease in $V/\sigma(R)$ to the outermost radii. This trend is reversed for velocity dispersion (Fig. \ref{fig:special_gals}, shown explicitly in the velocity dispersion map), and is unique to embedded disc galaxies. The velocity dispersion of an embedded disc galaxy is enhanced along the minor rotation axis, and decreases radially along the major rotation axis. The distinctive structure in the embedded disc velocity dispersion profile is highlighted against the velocity dispersion profiles of morphologically-standard galaxies shown in Figs. \ref{fig:morphology_gals} and \ref{fig:special_gals}. The three embedded disc galaxies SAMI 56140, SAMI 136841, and SAMI 347432 each have an $I_\text{ED}$ above, or just below, $I_\text{ED}=4$. 

One galaxy in the sample demonstrates a rare, double-peaked velocity dispersion in its two-dimensional velocity dispersion map. Such a galaxy has been termed a 2$\sigma$ galaxy \citep{Krajnovic11, Alabi15}. The 2$\sigma$ galaxy in our sample, SAMI 239560, is identified through a characteristic spin-ellipticity radial track of a slight increase in the local spin parameter with increasing ellipticity, followed by a dramatic decrease in spin corresponding to a marginal loss in ellipticity with radius. The velocity dispersion profile of this galaxy indicates a radially increasing velocity dispersion over the probed radii, as in Fig. \ref{fig:special_gals}. Conversely, the velocity profile of the 2$\sigma$ galaxy decreases beyond $\simeq1R_\text{e}$. The ratio $V/\sigma(R)$ monotonically decreases with radius, though the decrease is minor ($\sim0.2$). The galaxy SAMI 99430 is also noted as a potential $2\sigma$ galaxy based on its spin-ellipticity profile (Fig. \ref{fig:tracks}), although the velocity dispersion map is not consistent with that of a 2$\sigma$ galaxy. Importantly, the $I_\text{ED}$ of the 2$\sigma$ galaxy SAMI 239560 is greater than other morpho-dynamical types, with $I_\text{ED}=10.9$. The track segment length of SAMI 239560 is found to be $\overline{L}_{\text{2}\sigma}=0.021$; how representative this value is of 2$\sigma$ galaxies in general is inconclusive without further data from more 2$\sigma$ galaxies. 

Barred spiral galaxies exhibit a distinct spin-ellipticity radial relation, reminiscent of that of embedded discs, but without the decreasing spin with radius. The ellipticity of barred spirals decreases at the outer radii. The most noticeable variation in the spin-ellipticity radial tracks of barred spiral galaxies (Fig. \ref{fig:tracks}) is the difference in local spin at $R_\text{min}$ and $R_\text{max}$. Barred spirals have the greatest track segment length of all the morpho-dynamical types, with a median value of $\overline{L}_\text{eSB}=0.053$, and thus have radial points that are well-spread in spin-ellipticity parameter space. Similar to that of regular early- and late-type spiral galaxies shown in Fig. \ref{fig:morphology_gals}, the $V/\sigma(R)$ profile for barred spiral galaxies increases with radius. In the sample, all barred spiral galaxies were identified as early spirals through visual inspection.

Within the spin-ellipticity radial tracks of each morpho-dynamical type, minor variation is apparent. Visual classification of galaxy morphology for those galaxies in the GAMA survey was initially performed on SDSS images \citep{Cortese19}, whilst we use higher-resolution HSC and KiDS data for ellipse profile fitting. Using the HSC and KiDS images, visual inspection indicates 3/107 of E, 17/169 of S0, 4/133 of eSp, 4/50 of lSp, and 2/10 of eSB galaxies did not demonstrate the spin-ellipticity radial track typical of their corresponding morpho-dynamical type. The catalogue ID, SDSS classification, HSC/KiDS classification, and the spin-ellipticity radial track classification for each of the outlying galaxies is presented in Table \ref{tab:outliers} in Appendix \ref{sec:appOutlier}.


\section{Discussion}\label{sec:discussion}

In this work we combine photometry and stellar kinematic analysis in a qualitative morpho-dynamical approach to produce spin-ellipticity radial tracks for each galaxy in the SAMI Galaxy Survey internal data release v0.11 sample with stellar kinematics out to 1.5$R_\text{e}$. We quantify the spin-ellipticity radial tracks by calculating the track segment length, the mean spacing between adjacent radial points in spin-ellipticity parameter space. We relate the features seen in the spin-ellipticity radial tracks to the physical structures and dynamical properties of galaxies of all morphologies, particularly in identifying unusual structures such as embedded and counter-rotating discs. These properties, whilst difficult to interpret in unilateral studies of photometry or stellar kinematics alone, are readily identified in the spin-ellipticity radial tracks. We describe the ``rules of behaviour'' for each morpho-kinematic type below. We note that inclination will influence the value and possibly the appearance of spin-ellipticity radial tracks as both the apparent ellipticity and spin values are higher when seen edge-on than when seen at other inclination angles.

\subsection{Elliptical Galaxies}\label{ssec:elliptical}
The spin-ellipticity radial tracks of four typical elliptical galaxies (SAMI 54202, SAMI 136605, SAMI 422406, and SAMI 505765) are shown in Fig. \ref{fig:tracks}. SAMI 136605 (Fig. \ref{fig:tracks}, diamond track) and SAMI 422406 (Fig. \ref{fig:tracks}, square track) display little net rotation, and are primarily pressure-supported. Little variation in the velocity profile results in a similar trend in the local spin, irrespective of velocity dispersion. This relation consequently explains the slightly extended spin-ellipticity radial tracks seen for elliptical galaxies with a larger variation in the velocity profile, e.g. SAMI 54202 and SAMI 505765. As expected, a smaller difference between $V/\sigma(R_\text{max})$ and $V/\sigma(R_\text{min})$ corresponds to a reduced spin-ellipticity radial track segment length, whereas a larger difference between $V/\sigma(R_\text{max})$ and $V/\sigma(R_\text{min})$ due to increased recession velocity corresponds to an extended track segment length. This correlation is unsurprising given the similar behaviour of $V/\sigma(R)$ and $\lambda(R)$. A reduction in the difference between $V/\sigma(R)$ at maximum and minimum radius is seen as a reduced difference in $\lambda(R_\text{max})$ and $\lambda(R_\text{min})$, restricting the spin domain of the track on the spin-ellipticity radial plot. 

Constraining the ellipticity domain, thus ensuring a short track segment length, is the weak trend of ellipticity on mean radius. The ellipticity of the elliptical galaxies in the sample show only minor variation, indicating a smooth internal substructure. Consistent with the findings of \citet{Emsellem07} in the SAURON Survey, elliptical galaxies in the SAMI sample typically have low ellipticity values of $\varepsilon(R)\lesssim0.3$; the spin-ellipticity radial tracks of elliptical galaxies are thus concentrated in the bottom left corner of the spin-ellipticity radial plot. Elliptical galaxies with minimal rotation produce a spin-ellipticity radial track concentrated about a point in spin-ellipticity space (i.e. a very short track segment length), whereas elliptical galaxies with an increased net rotation produce an extended track segment length within the $\varepsilon(R)\lesssim0.3$ and $\lambda(R)\lesssim0.4$ bounds, due to a greater difference in primarily spin with increasing radius. SAMI 238500 (Fig. \ref{fig:morphology_gals}) clearly demonstrates the spin-ellipticity radial track of an elliptical galaxy with an increased net rotation, compared to the spin-ellipticity radial track of the slow rotator SAMI 136605 (diamond track in Fig. \ref{fig:tracks}, Elliptical panel). 

\subsection{Lenticular Galaxies}\label{ssec:lenticular}
Lenticular galaxies typically produce positively-sloped spin-ellipticity radial profiles (Fig. \ref{fig:tracks}): this applies to 156/169 of S0 galaxies in the sample. Deviations from this trend are likely due to structures in the images, such as bars, that reduce the ellipticity with increasing radius (see Sections \ref{sec:analysis} and \ref{ssec:spiral}). Noticeably, the sharp reversal in ellipticity at large radii, as seen in 5 of 169 galaxies (SAMI 84680, SAMI 98097, SAMI 227278, SAMI 238412, and SAMI 387715 in Appendix \ref{sec:app1p5}), corresponds to increasingly circularised isophotes. Comparing the sudden reversal in lenticular galaxy ellipticity to that displayed by a large proportion of early-spiral galaxies (Section \ref{ssec:spiral}) suggests sharp direction changes within a spin-ellipticity radial track may be related to changes in the radial photometric dominance of various structural components (e.g. bulge, disc, and/or halos). In the five lenticular galaxies with a notable reversal in ellipticity at large radii, a bar structure is present in the HSC and KiDS image of SAMI 84680 and SAMI 387715 respectively (Appendix \ref{sec:app1p5}). Both SAMI 227278 and SAMI 238412 have a monotonically increasing velocity profile an monotonically decreasing velocity dispersion profile (Appendix \ref{sec:app1p5}), however the KiDS images for both suggests SAMI 227278 and SAMI 238412 contain an embedded disc structure. The spin-ellipticity radial tracks for SAMI 227278 and SAMI 238412 indicate a sudden reversal in ellipticity consistent with that exhibited by embedded disc galaxies, however the spin-ellipticity radial track of SAMI 238412 does not have decrease in spin with increasing radius. The lack of radially-decreasing spin in the spin-ellipticity radial tracks for SAMI 238412 may be explained by the embedded disc structure extending beyond the probed radii. Critically, if an embedded disc galaxy is not probed to radii such that spin (and hence $V/\sigma$) begins to radially decrease, $(V/sigma)_\text{max}$ and $V/\sigma(R_\text{max})$ will coincide, forcing the $I_\text{ED}$ to 0. This is true for both SAMI 227278 and SAMI 238412 (i.e. $I_\text{ED}=0$).

The rotation of lenticular galaxies increases while the velocity dispersion decreases with radius, with the notable exceptions of SAMI 321035, SAMI 375564, and SAMI 534763. SAMI 321035 and SAMI 534763 have a radially-increasing velocity dispersion driven by a small number of outlying spaxels at outer radii, whereas SAMI 375546 has a definite radial increase in velocity dispersion, the cause of which is unknown. The majority of embedded disc galaxies (Section \ref{ssec:embed}) were classified as lenticular galaxies, but are readily distinguished through their spin-ellipticity radial tracks, as seen in Fig. \ref{fig:tracks} (also see Section \ref{ssec:embed}). 

\subsection{Spiral Galaxies}\label{ssec:spiral}
Similar to lenticular galaxies, 77 out of 133 of the early spiral galaxies demonstrate a positively-sloped spin-ellipticity radial track (Fig. \ref{fig:tracks}, early spiral Type I panel). The remaining 56 do not show strict adherence to a radially increasing ellipticity trend, thus giving a variety of shapes in the spin-ellipticity radial profiles (Fig. \ref{fig:tracks}, early spiral Type II panel).

The median track segment length of early spirals is greater than lenticular galaxies: $\overline{L}_\text{eSp}=0.038$, compared to $\overline{L}_\text{S0}=0.027$ (Fig. \ref{fig:tracklength}). Local spin consistently increases with radius, with the exception of local spin at the maximum radius of six early spiral galaxies, which decreases slightly (SAMI 40164, SAMI 178197, SAMI 202398, SAMI 347470, SAMI 485924, and SAMI 618108). Of these six, SAMI 40164 shows clear substructures indicative of a recent merger on the KiDS image, and SAMI 178197 is a 2$\sigma$ galaxy candidate from its velocity dispersion map. 

Increased fluctuations in the ellipticity and spin of early spirals occurs particularly at large radii, when compared to lenticular galaxies. These fluctuations are consistent with isophotal ellipses being fit through the spiral arms. Importantly, early spirals of Type II do not follow the positively-sloped spin-ellipticity trend for outer radii (Fig. \ref{fig:tracks}, early spiral Type II panel), but have spin-ellipticity radial tracks that may be naturally `split' into two segments with a positive-gradient segment and a negative-gradient segment. The first segment is indicated by radially-increasing ellipticity, and the second radially-decreasing ellipticity: this pattern is clearly exemplified by SAMI 492384 (Fig. \ref{fig:morphology_gals}). The two-part ellipticity profile of many early spirals (seen in Fig. \ref{fig:tracks}, Early Spiral II panel) further emphasises the importance of the differing bulge-disc structures, with the `turning point' in the spin-ellipticity radial profile indicating the radial extent of a highly-elliptical bulge structure. Spiral galaxies of Type I do not have a turning point in spin-ellipticity space. The lack of a turning point in the spin-ellipticity tracks of Type I early spirals is explained by the ellipticity of the bulge not differing significantly from the ellipticity of the disc.

Spiral galaxies of Type II  do have a turning point in spin-ellipticity space, indicating a transition from elliptical bulge to disc domination of the light profile. Prior to the turning point, dramatic increases in ellipticity indicates an elliptical bulge, or weak bar structure. The ensuing ellipticity decrease is consequently a return to the increasingly-circular isophote regime, as a function of radius. The bulge-disc domination interpretation may naturally be extended to explain barred spiral galaxy profiles, whereby the radial variation in ellipticity is greatly exaggerated compared to early spirals, and the turning point in spin-ellipticity space occurring at a median value of 0.88$R_\text{e}$ (Section \ref{ssec:bar}). Variation seen in the ellipticity beyond the turning point between different galaxies is due to detailed substructures (i.e. the spiral arms), causing rapid variation in the ellipse-fitting used to model the outer radii of the galaxy. Ellipticity variation caused by substructures (such as spiral arms) is supported by the lack of variation in the majority of lenticular galaxies within the sample, which have a smooth radial-luminosity profile extending to the outermost radii. 

It is unclear whether the early spiral class is truly bimodal, however one hypothesis for the occurrence of the two qualitative trends in the spin-ellipticity radial tracks is the presence of a less elliptical outer section than inner section for Type II such that $\varepsilon_\text{outer}<\varepsilon_\text{inner}$, and $\varepsilon_\text{outer}>\varepsilon_\text{inner}$ for Type I. 

In contrast to early-type spirals, a large proportion of late spirals do not necessarily demonstrate positively-sloped spin-ellipticity radial tracks. As shown in Fig. \ref{fig:tracks}, late-type spirals span a range of slopes, including some with a negative spin-ellipticity slope (e.g. SAMI 511921, diamond points in Fig. \ref{fig:tracks} Late Spiral panel), and others with a near-vertical slope (e.g. SAMI 78791, circles in the ``Late Spiral'' panel of Fig. \ref{fig:tracks}). The median track segment length of late spirals is seen to be comparable to that of lenticulars, with $\overline{L}_\text{lSp}=0.028$ (Fig. \ref{fig:tracklength}). Applying the same bulge-disc interpretation of the spin-ellipticity radial profile of early spirals to late spirals, only eight from a total of sixty-four galaxies may be divided naturally into two segments, with the remainder following straight-line profiles with varying slopes. The lack of a large proportion of late spirals with sharp reversals in the spin-ellipticity radial tracks highlights the reduced prominence of bulges in late spiral galaxies. This interpretation is supported by the respective velocity dispersion maps, which are constant for galaxies with a straight-line spin-ellipticity radial track, compared to the prominence of increased velocity dispersion in the galactic centre in early spiral galaxies. The $V/\sigma(R)$ profiles of both early and late spirals are consistent with $V/\sigma(R)$ increasing uniformly with radius. 

Four spiral galaxies (SAMI 55245, SAMI 296742, SAMI 422289, and SAMI 543895) were found to possess no defined rotation. Inspecting the HSC image of SAMI 55245 reveals the galaxy to be a merger, with a large shell visible to the left of the galactic centre. SAMI 543895, from its velocity map, has no rotation at low radii, consistent with an unresolved kinematically decoupled core. Rotation is apparent in the outer radii of the velocity map, at about 2$R_\text{e}$. The cause of a lack of rotation in the remaining two anomalous spirals, SAMI 296742 and SAMI 422289, is unclear (both are shown in Appendix \ref{sec:app1p5}).

\subsection{Embedded Disc Galaxies}\label{ssec:embed}
The spin-ellipticity radial tracks of embedded disc galaxies produce a distinct counter-clockwise signature, the result of a radial decrease in ellipticity and an initial radial increase, followed by a radial decrease, in the local spin (Fig. \ref{fig:tracks}), a pattern consistent with the hypothesis of \citet{Graham17}. Three galaxies: SAMI 56140 (square track), SAMI 136841 (circle track), and SAMI 347432 (triangle track), shown in Fig. \ref{fig:tracks}, have an embedded disc structure determined through the spin-ellipticity radial tracks. A further four galaxies (SAMI 91697, SAMI 505341, SAMI 545903, and SAMI 618976) are identified as potential embedded disc candidates. SAMI 56140 has an $I_\text{ED}>4$ (Fig. \ref{fig:EDI_plot}), whereas SAMI 136841 and SAMI 347432 both have an $I_\text{ED}<4$. Consequently, the embedded disc structure of SAMI 136841 and SAMI 347432 would prove difficult to be identified from kinematics alone \citep[e.g.][]{Foster18}. The first turning point in the spin-ellipticity radial track of an embedded disc galaxy corresponds to a sudden reversal in ellipticity (similar to that of early spirals), whilst the second turning point in the spin-ellipticity radial track corresponds to a sudden reversal in local spin. Turning points that are well-spaced in radius provide the spin-ellipticity radial track greater curvature, e.g. SAMI 56140. In Fig. \ref{fig:tracklength}, the median track segment length of embedded disc galaxies, $\overline{L}_\text{ES}=0.034$, is seen to be greater than both elliptical and lenticular galaxies: the two morpho-dynamical types that embedded disc galaxies span. The high $\overline{L}_\text{ES}$ in embedded disc galaxies compared to elliptical and lenticular galaxies is a result of greater variation in $\varepsilon(R)$ and $\lambda(R)$ with radius.

Embedded disc galaxies have a radially dependent initial increase, with a subsequent decrease, in the $V/\sigma(R)$ profile, and increased velocity dispersion along the minor rotation axis compared to the major rotation axis \citep[i.e.][]{Arnold14, Foster18}. Of the three identified embedded disc galaxies, each has a $V/\sigma(R)$ profile consistent with that expected from an embedded disc, and strong evidence for increased velocity dispersion along the minor rotation axis. Of these galaxies, SAMI 56140 presents the clearest evidence of an embedded disc structure (Fig. \ref{fig:special_gals}), consistent with the findings of \citet{Foster18}. SAMI 347432 is also identified as an embedded disc galaxy by \citet{Foster18}, though its spin-ellipticity radial track has less defined curvature than that of SAMI 56140 (Fig. \ref{fig:tracks}). Difference in curvature in the spin-ellipticity radial track of embedded disc galaxies may be attributed to a difference in disc orientation relative to the bulge. In the case of SAMI 56140, the disc is seen through visual inspection to be misaligned with respect to the bulge. Viewing inclination is also expected to affect the spin-ellipticity radial track of a galaxy. A galaxy viewed from an edge-on perspective is expected to have higher spin and ellipticity values than when viewed face-on. A profile with the distinguishing embedded disc curvature occurring at greater radii (e.g. SAMI 347432) is indicative of a larger embedded disc component, compared to a galaxy with a profile having the distinct curvature occurring at lower radii (e.g. SAMI 56140). Thus, from the spin-ellipticity radial plot, the size of the embedded disc structure relative to the total galaxy size may be estimated. 

\subsection{2-Sigma Galaxies}\label{ssec:2sig}
From the SAMI sample, the double dispersion peaks in the two-dimensional velocity dispersion map of SAMI 239560 (Fig. \ref{fig:special_gals}) shows compelling evidence of a counter-rotating structure. This galaxy was initially identified through an anomalous spin-ellipticity radial profile, and confirmed by its velocity dispersion map and $V/\sigma(R)$ profile. The spin-ellipticity radial track demonstrates an initial increase in the local spin parameter with increasing ellipticity, followed by a sudden decrease in spin and marginal loss in ellipticity with radius. The track segment length of SAMI 239560 is short: $\overline{L}_{2\sigma}=0.021$. The spin-ellipticity radial track of SAMI 239560 is consistent within the probed radial range with the spin-ellipticity radial track of the 2-sigma galaxy NGC 4473, presented in \citet{Bellstedt17b} from the SLUGGS Survey. As in the second panel of the second row of Fig. \ref{fig:special_gals}, the dominant recessional velocity at outer radii is partially bordered by an oppositely-directed recessional velocity, hinting at the counter-rotating structure. A lack of opposing recessional velocities enclosing the inner recessional velocities at the outer measured radii is attributed to not having data at large enough radii. From Fig. \ref{fig:special_gals}, the data for SAMI 239560 extends to just beyond $\sim1.5R_\text{e}$, thus if the radius at which both counter-rotating discs contribute equally to the measured rotation were located beyond 1.5$R_\text{e}$, the differing recessional velocities would not be apparent in the velocity map. 

Using the spin-ellipticity radial profile of SAMI 239560 as an expected 2$\sigma$ galaxy signature, one other potential 2$\sigma$ galaxy, SAMI 99430, was identified (Fig. \ref{fig:tracks}). The velocity dispersion map of SAMI 239560 clearly exhibits minimum dispersion in the central regions of the galaxy, monotonically increasing with radius (Fig. \ref{fig:special_gals}), whereas SAMI 99340 does not display the same clear decreased velocity dispersion in the galactic centre. The velocity dispersion profile of SAMI 239560 is consistent with the 2$\sigma$ galaxy NGC 4473 \citep[e.g.][]{Krajnovic11, Alabi15}. Interestingly, the embedded disc index (Equation \ref{eq:ied}) was exceptionally high for SAMI 239560, $I_\text{ED}>10$ (Fig. \ref{fig:EDI_plot}). This is owing to the large difference in $(V/\sigma)_\text{max}$ and $V/\sigma(R_\text{max})$, as shown in the fifth panel of the second row of Fig. \ref{fig:special_gals}: the $V/\sigma(R)$ profile monotonically decreases with radius, thus $(V/\sigma)_\text{max}$ and $V/\sigma(R_\text{max})$ do not coincide. As $(V/\sigma)_\text{max}$ and $V/\sigma(R_\text{max})$ should not coincide for any 2$\sigma$ galaxy, 2$\sigma$ galaxies should have a non-zero $I_\text{ED}$. Indeed, SAMI 99430 has a small but non-zero $I_\text{ED}$ of 1.99.

\subsection{Barred Spiral Galaxies}\label{ssec:bar}
The spin-ellipticity radial tracks traced by barred spiral galaxies in the SAMI sample demonstrate a definite pattern (Fig. \ref{fig:tracks}). As with lenticular and early spiral galaxies, the spin-ellipticity radial tracks of all barred spiral galaxies show a clear transition from the bar/bulge to disc regimes, seen as a sudden reversal from radially increasing ellipticity to decreasing ellipticity. Prior to the change in the ellipticity-radial relation, barred spiral galaxies show, as a function of radius, either a dramatic increase in ellipticity and a modest increase in spin (e.g. SAMI 422286, diamond points in Fig. \ref{fig:tracks} Barred Spiral panel, $\Delta\varepsilon(R)\simeq0.5$, $\Delta\lambda(R)\simeq0.3$), or a dramatic increase in spin and a less-pronounced increase in ellipticity (e.g. SAMI 17314, square track in Fig. \ref{fig:tracks} Barred Spiral panel, $\Delta\varepsilon(R)\simeq0.15$, $\Delta\lambda(R)\simeq0.5$). The dramatic increase in either $\varepsilon(R)$ or $\lambda(R)$ is responsible for the very high median track segment length of barred spirals, $\overline{L}_\text{eSB}=0.053$ (Fig. \ref{fig:tracklength}). As was the case for spiral galaxies, the turning point in the spin-ellipticity radial track of barred spiral galaxies is caused by ellipse fitting to two components: the bar and the disc. Prior to the sudden ellipticity reversal, the rapid increase in ellipticity with radius of a barred spiral galaxy is an exaggerated version of a non-barred early-type spiral due to the high ellipticity of the bar structure (e.g. in particular SAMI 422286). As for spiral galaxies, the radial transition from bar to disc domination is determined by the turning point in the spin-ellipticity radial profile. The increase of spin with radius (seen as the vertical displacement of radial points in spin-ellipticity space) is fairly consistent both prior to and beyond the turning point in the spin-ellipticity radial track. Consequently, galaxies with a dramatic increase in spin prior to the reversal in ellipticity continue to have a dramatic increase in spin beyond the ellipticity reversal point (e.g. SAMI 17314, Fig. \ref{fig:tracks}). Crucially, the local spin increases with radius irrespective of ellipticity, which is also reflected as an increase in the $V/\sigma(R)$ ratio with radius.

\subsection{Differentiating Barred Spiral and Embedded Disc Galaxies}\label{ssec:BvsED}
Comparing the spin-ellipticity radial tracks of barred spiral and embedded disc galaxies, the distinction is subtle. The key difference is the number of turning points within the profile. Neglecting the outermost radial point, barred spiral galaxies have one turning point: where ellipticity suddenly decreases with radius. On the other hand, embedded disc galaxies have two turning points: the first where ellipticity reverses from radially increasing to radially decreasing, and the second, where spin dramatically decreases. In the particular case of SAMI 56140, these turning points are well-separated in radius, and are related to the three regimes of the velocity dispersion map. At small radii, an embedded disc galaxy behaves similar to a typical disc (e.g. S0) galaxy, with a radial decrease in velocity dispersion corresponding to the first turning point in the spin-ellipticity radial track, analogous to the turning point exhibited by early spiral galaxies. Unlike early spiral galaxies, the velocity dispersion for an embedded disc galaxy then increases at large radii, corresponding to a second turning point and a subsequent decline in spin. The definite local spin decrease seen at extreme radii in embedded disc galaxies is not seen in barred spiral galaxies. This is confirmed by the behaviour of $V/\sigma(R)$ for barred spiral and embedded disc galaxies.

\section{Conclusions}\label{sec:conclusions}
In this work, we study the use of spin-ellipticity radial tracks as a morpho-dynamical approach to classifying galaxies for 507 SAMI galaxies with stellar kinematics out to 1.5$R_\text{e}$. Our conclusions are presented below.
\begin{enumerate}
    \item We show that a bilateral approach combining photometry and stellar kinematics produces spin-ellipticity radial tracks that are particular to each morpho-dynamical type. The spin-ellipticity radial tracks highlight structural differences within a galaxy, in particular embedded discs and counter-rotating discs, which would otherwise not be seen in a unilateral study of photometry or kinematics alone. Consequently, seven morpho-dynamical types are identified: elliptical, lenticular, early spiral, late spiral, barred spiral, embedded disc, and 2-sigma galaxies. 
    \item The radial transition from bulge to disc domination of spirals, particularly barred spirals, is found from the sudden ellipticity reversal from radially-increasing to radially-decreasing ellipticity in the spin-ellipticity radial track. The lesser prominence of a bulge in late spirals compared to early spirals is seen through a lack of a sudden reversal in ellipticity in the spin-ellipticity radial track.
    \item The importance of probing the outer radii of galaxies in morphological and kinematic studies is highlighted by the features found at the galactic outskirts in the spin-ellipticity radial tracks, such as the radial transition from bulge domination to disc domination of the light profile.
\end{enumerate}
The use of spin-ellipticity radial tracks presents a powerful, information-dense, hybrid morpho-dynamical taxonomy method to both classify and confirm the structure of galaxies. This work emphasises the need to treat galaxies not as single points on the spin-ellipticity diagram, but instead explicitly probe radial variations.

\section*{Acknowledgements}

We would like to thank A. Graham for initial discussions that were the catalyst for using spin-ellipticity radial tracks. We also offer our thanks to A. Robotham for help with the {\sc ProFound} software tool.

This research was conducted by the Australian Research Council Centre of Excellence for All Sky Astrophysics in 3 Dimensions (ASTRO 3D), through project number CE170100013. 
The SAMI Galaxy Survey is based on observations made at the Anglo-Australian Telescope. The Sydney-AAO Multi-object Integral field spectrograph (SAMI) was developed jointly by the University of Sydney and the Australian Astronomical Observatory. The SAMI input catalogue is based on data taken from the Sloan Digital Sky Survey, the GAMA Survey and the VST ATLAS Survey.
The SAMI Galaxy Survey is supported by the Australian Research Council Centre of Excellence for All Sky Astrophysics in 3 Dimensions (ASTRO 3D), through project number CE170100013, the Australian Research Council Centre of Excellence for All-sky Astrophysics (CAASTRO), through project number CE110001020, and other participating institutions. The SAMI Galaxy Survey website is http://sami-survey.org/. 

JJB acknowledges support of an Australian Research Council Future Fellowship (FT180100231).
JBH is supported by an ARC Laureate Fellowship that funds Jesse van de Sande and an ARC Federation Fellowship that funded the SAMI prototype.
S.K.Y. acknowledges support from the Korean National Research Foundation (2017R1A2A1A05001116).
M.S.O. acknowledges the funding support from the Australian Research Council through a Future Fellowship (FT140100255).
SB acknowledges the funding support from the Australian Research Council through a Future Fellowship (FT140101166).
Support for AMM is provided by NASA through Hubble Fellowship grant \#HST-HF2-51377 awarded by the Space Telescope Science Institute, which is operated by the Association of Universities for Research in Astronomy, Inc., for NASA, under contract NAS5-26555.




\bibliographystyle{mnras}
\bibliography{SpinEllipticity.bbl} 


\newpage
\appendix

\renewcommand\thefigure{A.\arabic{figure}}
\section{Diagnostic Plots of 1.0$R_\text{e}$}\label{sec:app1p0}
\setcounter{figure}{0}
To justify the decision to limit our study to those galaxies with stellar kinematics out to at least 1.5$R_\text{e}$, we initially increase the sample size to include all galaxies with measured stellar kinematics out to 1$R_\text{e}$. This reduced radial range is often insufficient in identifying galaxies with defining features located at beyond the inner radii (e.g. spiral arms). Barred spiral galaxies with stellar kinematics out to only 1$R_\text{e}$ were particularly poorly-resolved in their spin-ellipticity radial tracks, as the bar is typically the dominant feature within the 0-1$R_\text{e}$ radial range. Those galaxies with radially-independent structure (i.e ellipticals) and structures located at small radii (i.e. embedded discs and counter-rotating discs) produced similar spin-ellipticity radial tracks as those galaxies with stellar kinematics out to 1.5$R_\text{e}$. We therefore limit our analysis and discussion to those galaxies with measured kinematic data out to 1.5$R_\text{e}$, however galaxies with interesting substructures able to be resolved with stellar kinematics out to 1$R_\text{e}$ are presented in diagnostic plots (Fig. \ref{fig:app_plot0}); these are embedded disc candidates and 2$\sigma$ galaxy candidates. Plotting $I_\text{ED}$ against $V/\sigma$ at $1R_\text{e}$ (as in Fig. \ref{fig:EDI_plot}) clearly identifies seven galaxies which are likely to possess an embedded disc (Fig. \ref{fig:EDI_plot}).

\begin{figure*}
    \centering
    \includegraphics[width=0.9\textwidth]{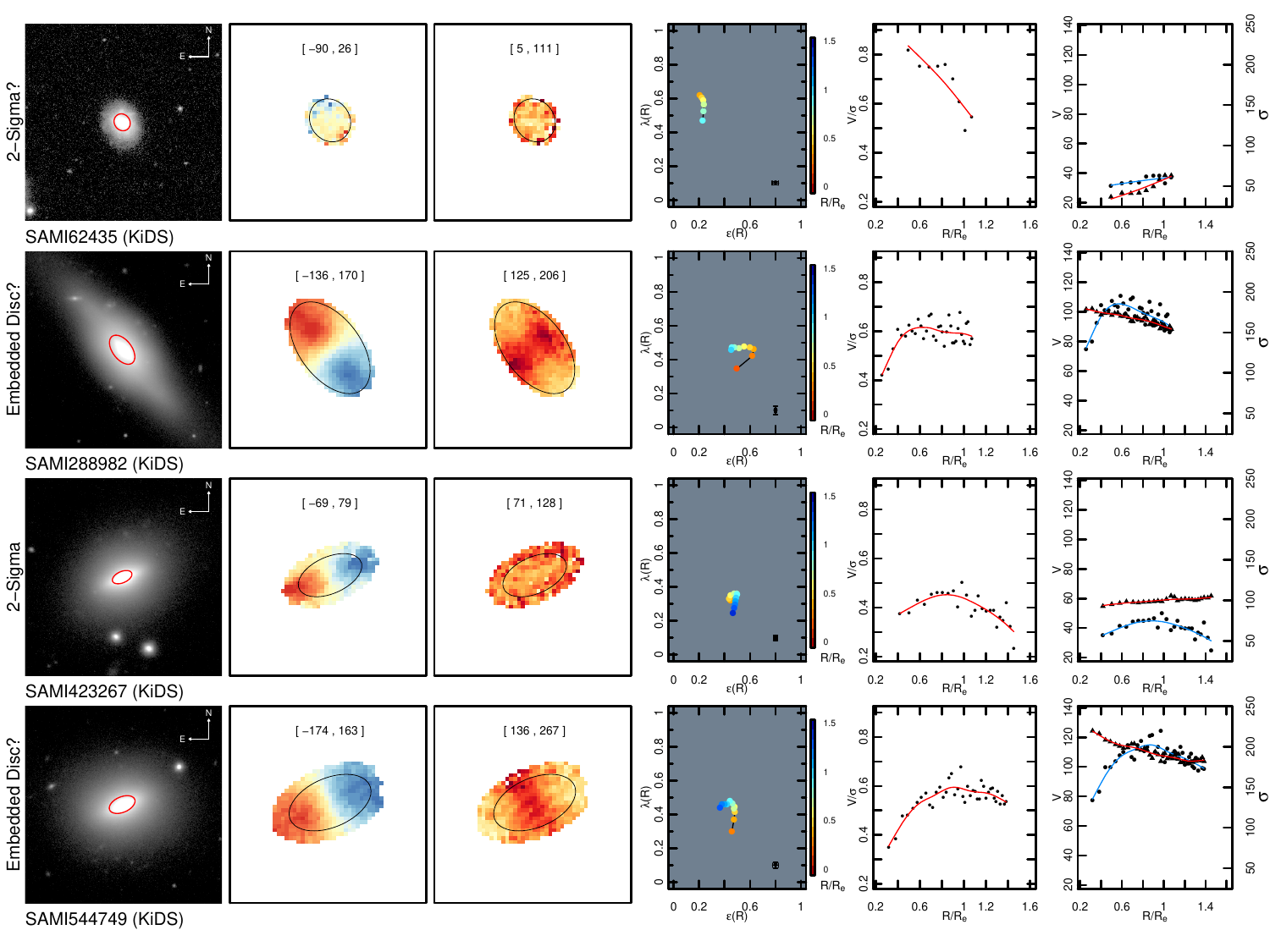}
    \caption{Diagnostic plots of select SAMI galaxies from the 1.0$R_\text{e}$ dataset. Columns as per Fig. \ref{fig:morphology_gals}, organised by catalogue ID, for SAMI 62453 to SAMI 544749. Note the maximum probed radii extends to only 1.5$R_\text{e}$.}
    \label{fig:app_plot0}
\end{figure*}

\newpage
\renewcommand\thefigure{B.\arabic{figure}}
\section{Diagnostic Plots of 1.5$R_\text{e}$ Dataset}\label{sec:app1p5}
\setcounter{figure}{0}
The diagnostic plots for all galaxies from the 1.5$R_\text{e}$ dataset discussed in the paper are presented for reference. Here, `NR' indicates a non-rotator, and a `?' an unconfirmed candidate for a specific morpho-dynamical type. The diagnostic plots and classification for all 507 galaxies in the study may be found in the supplementary material.

\begin{figure*}
    \centering
    \includegraphics[width=0.9\textwidth]{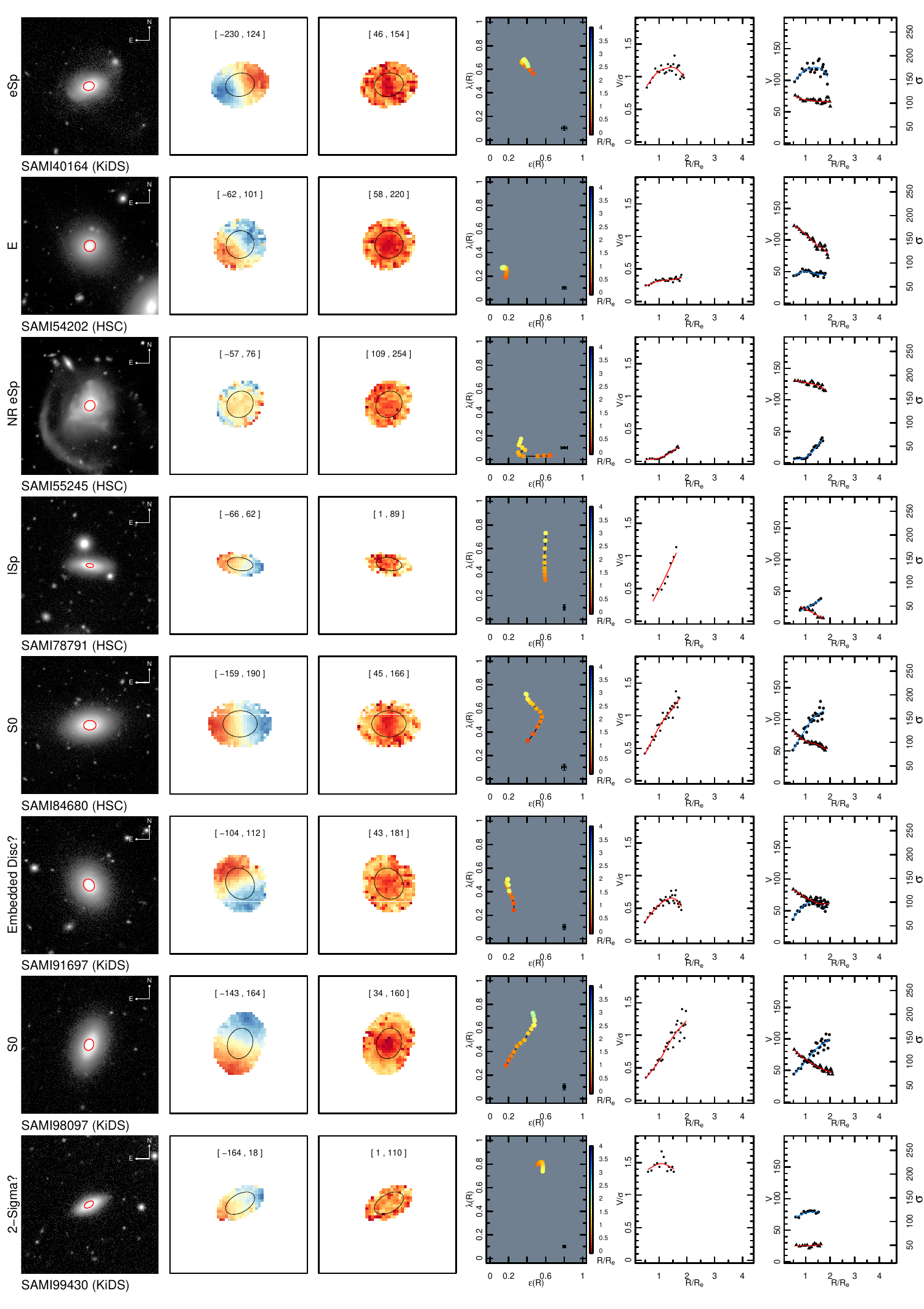}
    \caption{Diagnostic plots of SAMI galaxies discussed in the paper. Columns as per Fig. \ref{fig:morphology_gals}, organised by catalogue ID, for SAMI 40164 to SAMI 99430. Note the radial range extends from 0.0$R_\text{e}$ to 4.0$R_\text{e}$.}
    \label{fig:app_plot1}
\end{figure*}

\begin{figure*}
    \centering
    \includegraphics[width=0.9\textwidth]{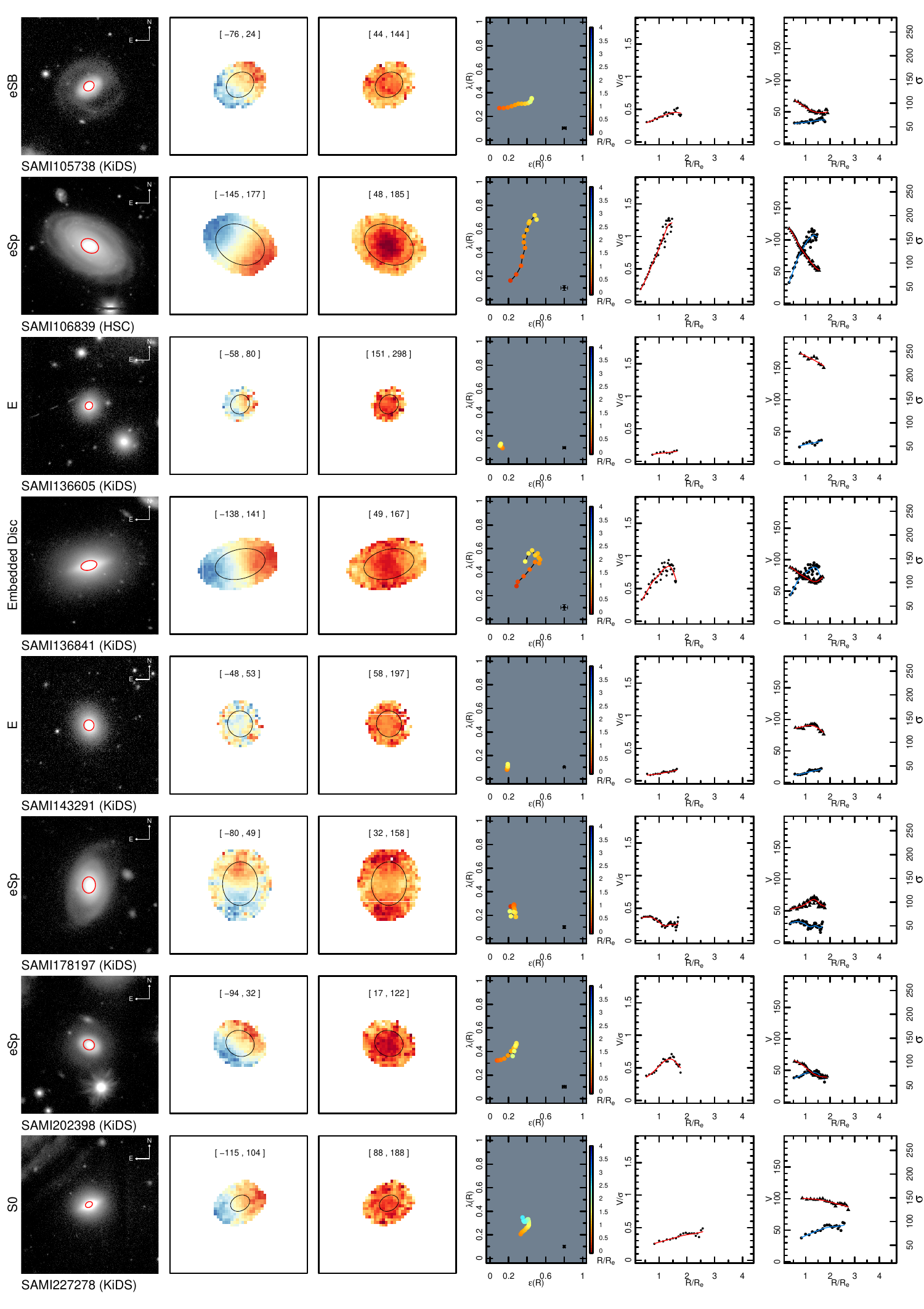}
     \caption{Diagnostic plots of SAMI galaxies discussed in the paper. Columns as per Fig. \ref{fig:morphology_gals}, organised by catalogue ID, for SAMI 105738 to SAMI 227278. Note the radial range extends from 0.0$R_\text{e}$ to 4.0$R_\text{e}$.}
    \label{fig:app_plot2}
\end{figure*}

\begin{figure*}
    \centering
    \includegraphics[width=0.9\textwidth]{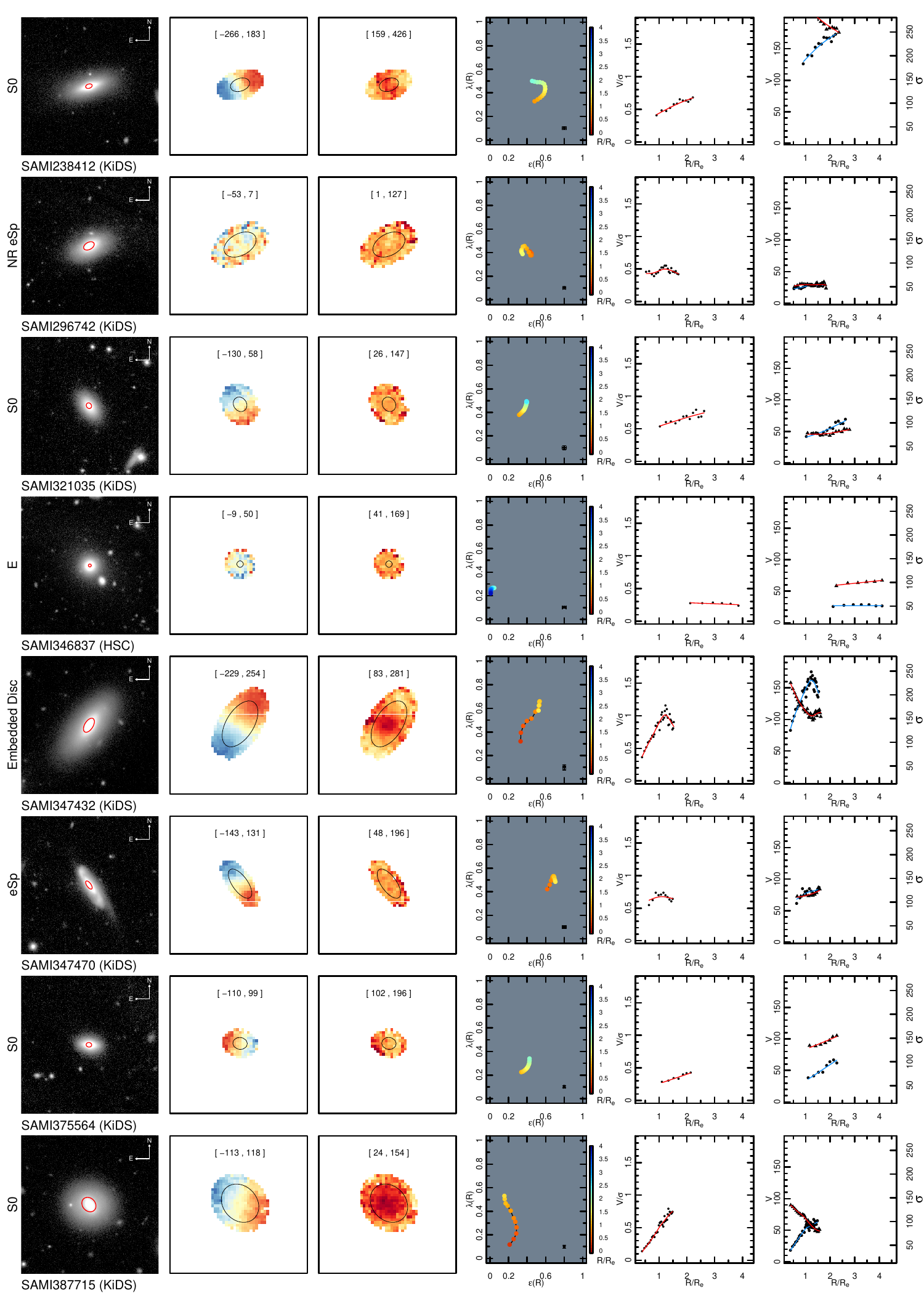}
     \caption{Diagnostic plots of SAMI galaxies discussed in the paper. Columns as per Fig. \ref{fig:morphology_gals}, organised by catalogue ID, for SAMI 238412 to SAMI 387715. Note the radial range extends from 0.0$R_\text{e}$ to 4.0$R_\text{e}$.}
    \label{fig:app_plot3}
\end{figure*}

\begin{figure*}
    \centering
    \includegraphics[width=0.9\textwidth]{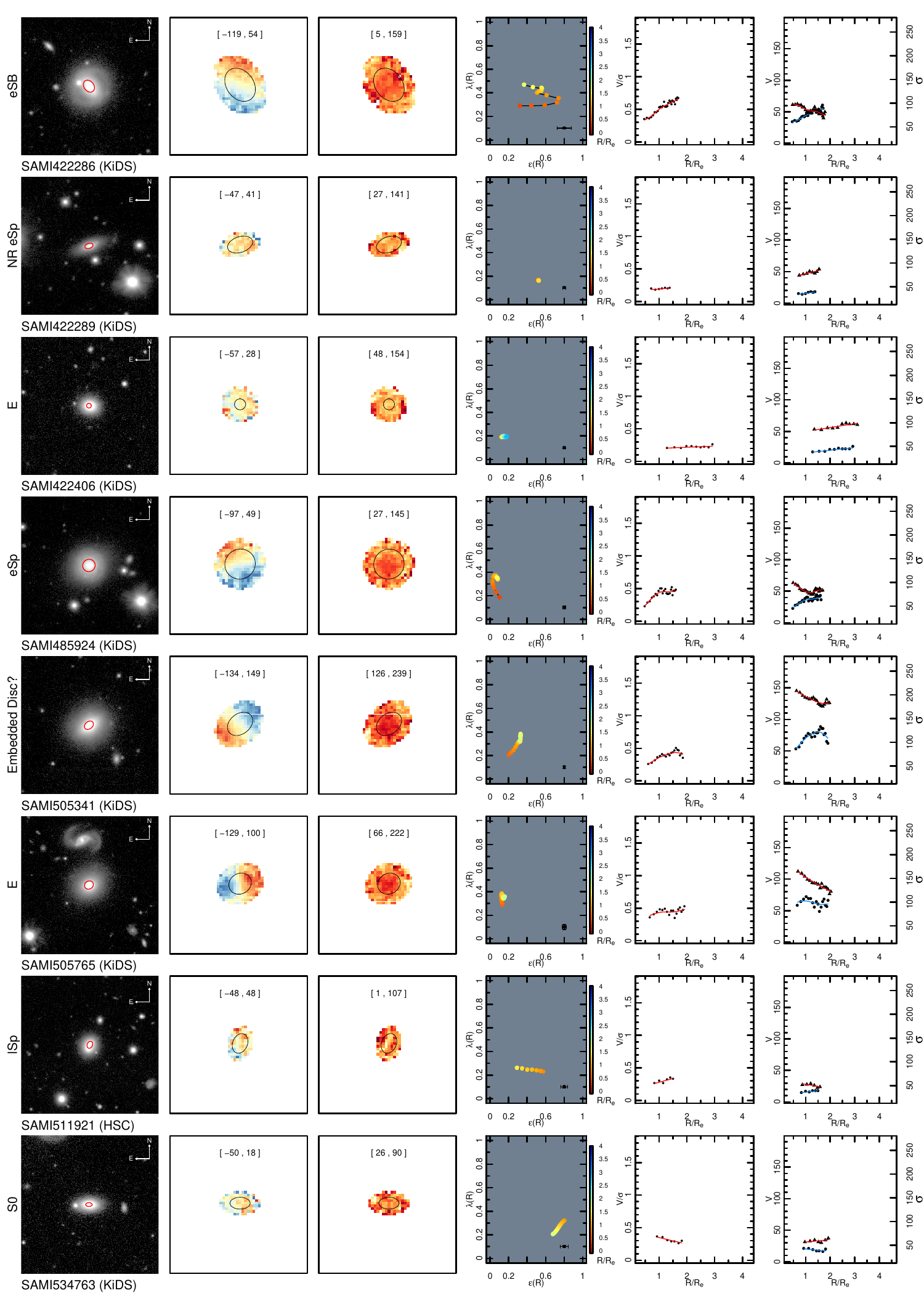}
     \caption{Diagnostic plots of SAMI galaxies discussed in the paper. Columns as per Fig. \ref{fig:morphology_gals}, organised by catalogue ID, for SAMI 422286 to SAMI 534763. Note the radial range extends from 0.0$R_\text{e}$ to 4.0$R_\text{e}$.}
    \label{fig:app_plot4}
\end{figure*}

\begin{figure*}
    \centering
    \includegraphics[width=0.9\textwidth]{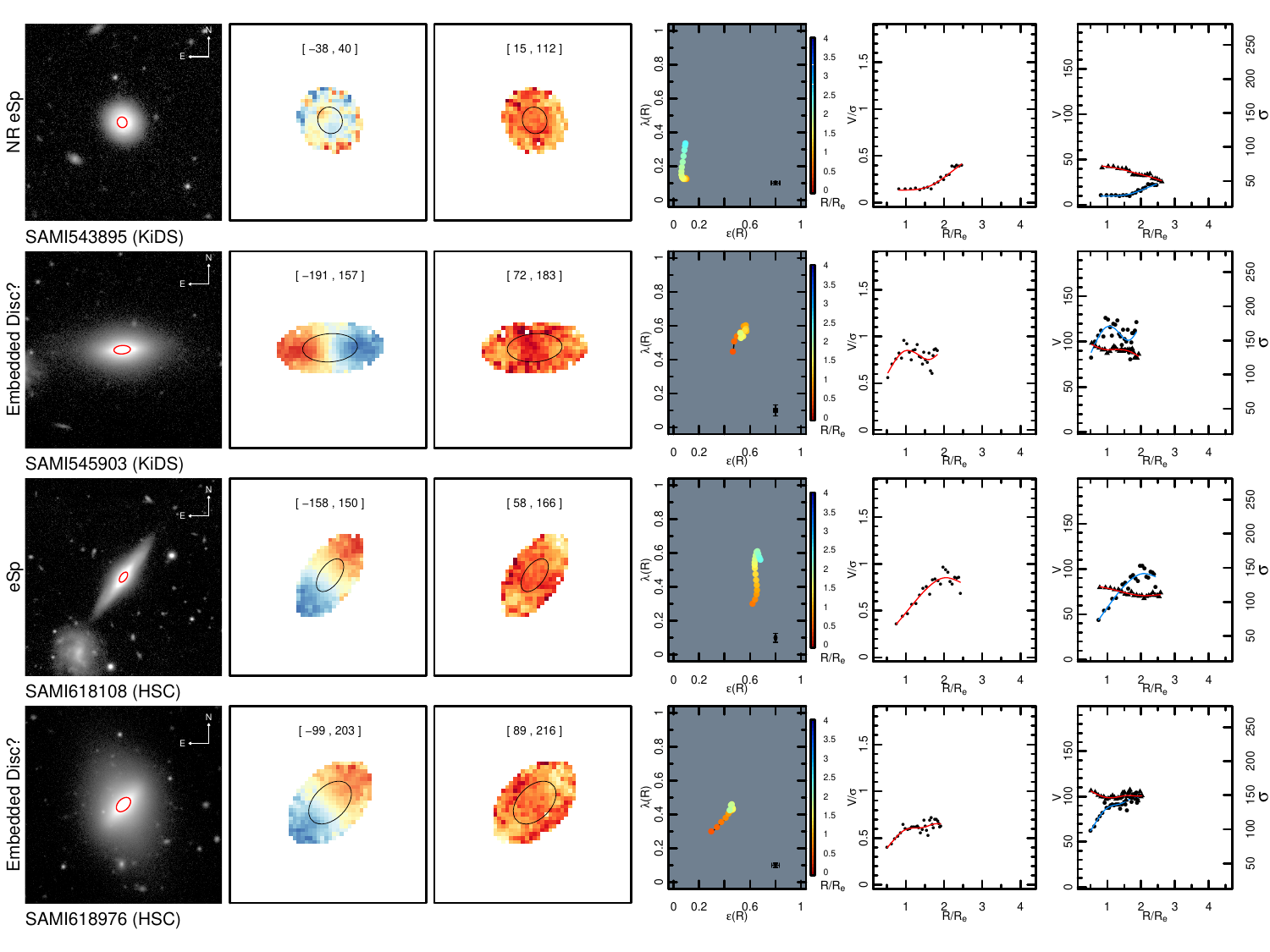}
     \caption{Diagnostic plots of SAMI galaxies discussed in the paper. Columns as per Fig. \ref{fig:morphology_gals}, organised by catalogue ID, for SAMI 543895 to SAMI 618976. Note the radial range extends from 0.0$R_\text{e}$ to 4.0$R_\text{e}$.}
    \label{fig:app_plot5}
\end{figure*}

\renewcommand\thefigure{C.\arabic{figure}}
\section{Galaxies with Outlying Spin-Ellipticity Radial Tracks}\label{sec:appOutlier}
\setcounter{figure}{0}
Galaxies with atypical spin-ellipticity radial tracks for their morpho-dynamical class, as described in Section \ref{ssec:sertracks}, are presented in Table \ref{tab:outliers}.
\begin{table}
    \centering
    \begin{tabular}{c|c|c|c}
        SAMI ID & SDSS & HSC/KiDS & Spin-Ellipticity \\
        \hline
         41354 & lSp & lSp  & ? \\
         48506 & lSp & lSp & ? \\
         93023 & eSp & eSp & ? \\
         105738 & eSB & eSB & eSp/lSp \\
         136616 & S0 & S0 & ? \\
         136688 & eSp & eSp & ? \\
         145073 & lSp & ? & ? \\
         184415 & lSp & lSp & eSp \\
         202399 & S0 & S0 & E/S0 \\
         220499 & S0 & S0 & eSB \\
         271387 & S0 & S0 & ? \\
         278840 & S0 & S0 & lSp \\
         323566 & E & E & E/S0 \\
         375615 & E & E & lSp \\
         382505 & E & E & eSp \\
         485924 & eSp & eSp & E \\
         491508 & S0 & S0 & eSp \\
         548946 & eSB & eSB & eSp \\
         570207 & S0 & S0 & ? \\
         574617 & eSp & eSp & lSp \\
    \end{tabular}
    \caption{Classifications of SAMI galaxies with atypical spin-ellipticity radial tracks. The original classification is presented in the SDSS column, followed by the classification from higher resolution HSC or KiDS data (Section \ref{ssec:photoprofile}). The classification from the spin-ellipticity radial tracks is given in column four. If classification is not possible in any column, a `?' is recorded.}
    \label{tab:outliers}
\end{table}

%
%
%

\bsp	
\label{lastpage}
\end{document}


\maketitle
    
    \begin{longtable}[c]{||c|c|c|c|c||}
    \caption{Classification of SAMI Galaxies with stellar kinematics out to at least 1.5$R_\text{e}$.}\label{tab:t1}
        \hline
        \multicolumn{5}{| c |}{Begin of Table \ref{tab:t1}}\\
        \hline
        Catalogue ID & SDSS & HSC/KiDS & Spin-Ellipticity & Kinematic Features\\
        \hline
        \endfirsthead
        
        \hline
        \multicolumn{5}{| c |}{Continuation of Table \ref{tab:t1}}\\
        \hline
        Catalogue ID & SDSS & HSC/KiDS & Spin-Ellipticity & Kinematic Features\\
        \hline
        \endhead
        
        \hline
        \endfoot
        
        \hline
        \multicolumn{5}{| c |}{End of Table \ref{tab:t1}}\\
        \hline\hline
        \endlastfoot
            7289 & E/S0 & E/S0 & E & - \\
            7841 & S0/eS & S0/eS & S0 & - \\
            7946 & eS & eS & eS & - \\
            8865 & E & E & E & - \\
            9062 & S0 & S0 & S0 & - \\
            14552 & S0 & S0 & S0 & - \\
            16022 & E/S0 & E/S0 & E & - \\
            16026 & eS/lS & eS/lS & eS/lS & - \\
            16317 & S0 & S0 & eS/S0 & - \\
            16429 & ? & ? & E? & - \\
            16525 & E & E & E & - \\
            16526 & E & E & E & - \\
            16926 & S0 & S0 & S0 & - \\
            17314 & eS & eSB & eSB & - \\
            22582 & E/S0 & E/S0 & E & - \\
            23070 & S0 & S0 & S0 & - \\
            23075 & S0/eS & S0/eS & S0 & - \\
            23082 & E & E & E & - \\
            23341 & E & E & E & - \\
            23565 & eS & eS & eS & - \\
            23619 & S0/eS & S0/eS & S0 & - \\
            28738 & lS & lS & eS/lS & - \\
            30786 & eS & eS & eS & - \\
            31912 & S0/eS & S0/eS & S0 & - \\
            39145 & eS/lS & eS/lS & lS & Rotational Shear \\
            39365 & eS & eS & eS & - \\
            39709 & S0/eS & S0/eS & S0 & - \\
            40164 & eS & eS & ? & Merger \\
            40197 & S0/eS & S0/eS & S0/eS & - \\
            40491 & E/S0 & E/S0 & S0 & - \\
            41180 & E & E & E & - \\
            41274 & E/S0 & E/S0 & E/S0 & - \\
            41302 & S0/eS & S0/eS & S0/eS & - \\
            41354 & lS & lS & lS & - \\
            47254 & E & E & E & - \\
            47258 & S0/eS & S0/eS & eS & - \\
            47293 & S0 & S0 & S0 & - \\
            47493 & lS & lS & lS & - \\
            47500 & lS & lS & lS & - \\
            48016 & eS/lS & eS/lS & eS & - \\
            48049 & S0/eS & S0/eS & eS & - \\
            48470 & E & E & E & - \\
            48506 & eS/lS & eS/lS & lS & - \\
            48681 & S0 & S0 & E/S0 & - \\
            49734 & S0 & S0 & S0 & - \\
            49784 & S0 & S0 & S0 & - \\
            54202 & E & E & E & - \\
            54957 & eS & eS & eS & - \\
            55245 & eS/lS & eS/lS & ? & Merger \\
            55537 & eS/lS & eS/lS & lS & - \\
            56140 & S0 & S0 & ES & Misaligned Disc \\
            56181 & E & E & E & - \\
            63276 & eS & eSB & eSB & - \\
            63780 & S0/eS & S0/eS & S0/eS & - \\
            64087 & eS/lS & eS/lS & lS & - \\
            65384 & eS & eS & eS & - \\
            65408 & E & E & E & - \\
            65410 & E/S0 & E/S0 & E & - \\
            69740 & ? & ? & ? & - \\
            69752 & eS & eS & eS & - \\
            69986 & eS & eS & ? & - \\
            70532 & S0/eS & S0/eS & S0 & - \\
            77227 & eS & eS & eS & - \\
            77710 & S0/eS & S0/eS & eS & - \\
            78355 & S0/eS & S0/eS & eS & - \\
            78530 & E/S0 & E/S0 & S0 & - \\
            78634 & S0 & S0 & S0 & - \\
            78791 & lS & lS & lS & - \\
            79693 & S0/eS & S0/eS & eS & - \\
            79711 & S0 & S0 & S0 & - \\
            79750 & E & E & E & - \\
            79810 & E & E & E & - \\
            79866 & E & E & E & - \\
            84106 & E/S0 & E/S0 & S0 & Rotational Shear \\
            84209 & E & E & E & - \\
            84680 & S0 & S0 & ? & - \\
            85239 & S0/eS & S0/eS & S0/eS & - \\
            85280 & eS & eS & eS & - \\
            85507 & lS & lS & 2-Sigma & - \\
            85667 & E & E & E & - \\
            91545 & E & E & E & - \\
            91568 & E/S0 & E/S0 & E/S0 & - \\
            91686 & S0 & S0 & S0 & - \\
            91689 & E & E & E & - \\
            91697 & E/S0 & E/S0 & ? & - \\
            91926 & S0/eS & S0/eS & eS & - \\
            91959 & S0/eS & S0/eS & S0/eS & - \\
            91996 & S0/eS & S0/eS & S0/eS & - \\
            91999 & ? & ? & ? & - \\
            92770 & ? & ? & ? & - \\
            92944 & lS & lS & lS & - \\
            93020 & S0/eS & S0/eS & S0/eS & - \\
            93023 & eS & eS & ? & - \\
            98097 & S0 & S0 & S0 & - \\
            99393 & eS & eS & eS &  \\
            99430 & lS & lS & 2-Sigma & - \\
            99742 & S0/eS & S0/eS & S0 & - \\
            100162 & lS & lS & lS & - \\
            105420 & eS & eS & eS & - \\
            105447 & S0/eS & S0/eS & S0 & - \\
            105519 & S0 & S0 & S0 & - \\
            105738 & S0 & eSB & eSB & - \\
            106047 & S0/eS & S0/eS & S0/eS & - \\
            106549 & E/S0 & E/S0 & E & - \\
            106634 & eS/lS & eS/lS & eS & - \\
            106810 & S0/eS & S0/eS & S0/eS & - \\
            106839 & S0 & eS & eS & Secular \\
            107454 & ? & ? & E & NR \\
            107503 & eS/lS & eS/lS & eS & Rotational Shear \\
            107824 & eS & eS & eS & - \\
            136232 & S0/eS & S0/eS & S0/eS & - \\
            136605 & E & E & E & - \\
            136616 & S0 & S0 & ? & - \\
            136624 & S0 & S0 & S0 & - \\
            136688 & eS & eS & eS & - \\
            136841 & S0 & S0 & Embedded Disc & - \\
            136871 & E/S0 & E/S0 & E & - \\
            136880 & S0/eS & S0/eS & S0/eS & - \\
            137809 & eS & eS & lS & - \\
            137841 & S0 & S0 & S0 & - \\
            138381 & lS & lS & lS & - \\
            143227 & S0/eS & S0/eS & S0/eS & - \\
            143291 & E & E & E & NR \\
            143605 & eS/lS & eS/lS & eS & - \\
            143736 & S0/eS & S0/eS & eS & - \\
            143746 & S0/eS & S0/eS & S0 & - \\
            143918 & S0 & S0 & S0 & - \\
            144051 & eS & eSB & eSB & - \\
            144197 & lS & lS & lS & - \\
            144320 & ? & ? & ? & KDC? \\
            144402 & eS/lS & Merger & ? & - \\
            144498 & S0 & S0 & S0 & - \\
            144640 & S0 & Merger? & S0 & - \\
            145073 & lS & lS & lS & - \\
            145729 & S0/eS & S0/eS & eS & - \\
            177010 & S0/eS & S0/eS & S0 & - \\
            177143 & eS & eS & eS & - \\
            177969 & E & E & E & NR \\
            178197 & eS & eS & ? & Merger, 2-sigma \\
            179584 & eS/lS & eS/lS & lS & - \\
            184046 & E & E & E & - \\
            184122 & eS & eS & eS & - \\
            184202 & S0 & S0 & E & NR \\
            184415 & lS & lS & lS & - \\
            184689 & S0/eS & S0/eS & S0 & - \\
            185085 & S0 & S0 & S0 & - \\
            185224 & S0 & S0 & S0 & - \\
            185365 & E/S0 & E/S0 & E & Prolate Rotator \\
            185567 & E/S0 & E/S0 & E & - \\
            186380 & lS & lS & lS & - \\
            186470 & S0 & S0 & S0 & - \\
            186573 & S0/eS & S0/eS & eS & - \\
            186803 & eS/lS & eS/lS & lS & - \\
            186470 & S0 & S0 & S0 & - \\
            186573 & S0/eS & S0/eS & eS & - \\
            186803 & eS/lS & eS/lS & ? & - \\
            197866 & eS & eS & ? & - \\
            198343 & S0 & S0 & S0 & - \\
            202360 & eS & eS & eS & NR \\
            202398 & eS & eS & eS & Prolate Rotator \\
            202399 & S0 & S0 & S0 & - \\
            202636 & lS & lS & lS & - \\
            203114 & eS & eS & eS & - \\
            203140 & S0 & S0 & S0 & - \\
            203833 & E & E & E & - \\
            204093 & S0 & S0 & S0 & - \\
            204490 & eS & eS & eS & - \\
            205020 & S0/eS & S0/eS & S0/eS & - \\
            205089 & S0 & S0 & S0 & - \\
            208876 & S0 & S0 & S0 & - \\
            209185 & E/S0 & S0/eS & S0/eS & - \\
            210228 & S0/eS & S0/eS & S0 & - \\
            210375 & S0 & S0 & S0 & - \\
            210467 & E & E & E & - \\
            210611 & eS & eS & eS & - \\
            214211 & E/S0 & E/S0 & E & - \\
            215292 & lS & lS & lS & - \\
            216396 & S0 & S0 & S0 & - \\
            218681 & lS & lS & lS & - \\
            219992 & S0 & S0 & S0 & - \\
            220217 & eS/lS & eS/lS & eS/lS & - \\
            220328 & S0 & S0 & S0 & - \\
            220374 & eS/lS & eS/lS & eS/lS & - \\
            220437 & E & E & E & - \\
            220499 & S0 & S0 & ? & - \\
            220502 & S0 & S0 & S0 & - \\
            220578 & lS & lS & lS & - \\
            227123 & eS & eS & eS & - \\
            227264 & S0/eS & S0/eS & S0/eS & - \\
            227278 & S0 & S0 & ? & - \\
            227905 & S0 & S0 & S0 & - \\
            228066 & S0/eS & S0/eS & S0 & - \\
            228428 & lS & lS & lS & - \\
            228564 & eS & eS & eS & - \\
            228767 & ? & ? & E & - \\
            229164 & E/S0 & E/S0 & E & - \\
            230421 & S0 & S0 & eS & - \\
            230683 & E & E & E & - \\
            230786 & E & E & E & - \\
            230787 & E & E & E & - \\
            230789 & E/S0 & E/S0 & E & - \\
            230793 & S0 & S0 & S0 & - \\
            230796 & ? & ? & E & - \\
            238204 & eS & eS & eS & - \\
            238276 & S0 & S0 & ? & - \\
            238282 & S0 & S0 & S0 & - \\
            238398 & ? & ? & E & - \\
            238412 & S0 & S0 & S0 & - \\
            238453 & E/S0 & E/S0 & E & - \\
            238500 & E & E & E &  -\\
            239249 & lS & lS & lS & - \\
            239292 & E/S0 & E/S0 & S0 & - \\
            239560 & eS/lS & eS/lS & 2-Sigma & - \\
            240352 & S0/eS & S0/eS & S0 & - \\
            250179 & ? & ? & E & - \\
            250192 & S0/eS & S0/eS & S0 & - \\
            250243 & E & E & E & - \\
            250466 & eS/lS & eS/lS & eS & - \\
            252057 & E/S0 & E/S0 & S0 & - \\
            271385 & eS & eS & eS & - \\
            271387 & S0 & S0 & ? & - \\
            271431 & S0/eS & S0/eS & S0/eS & - \\
            271455 & eS & eS & eS & - \\
            272190 & E & E & E & - \\
            272298 & eS & eS & eS & - \\
            272439 & S0 & S0 & S0 & - \\
            272488 & E/S0 & E/S0 & S0 & - \\
            272820 & E & E & E & - \\
            272990 & E/S0 & E/S0 & E/S0 & - \\
            278109 & S0/eS & S0/eS & eS & - \\
            278840 & E/S0 & E/S0 & S0 & - \\
            279148 & eS/lS & eS/lS & eS/lS & - \\
            279905 & E/S0 & E/S0 & S0 & - \\
            279910 & E/S0 & E/S0 & E & - \\
            287827 & S0/eS & S0/eS & S0/eS & - \\
            288679 & S0 & S0 & S0 & - \\
            288803 & E & E & E & - \\
            289102 & S0 & S0 & S0 & - \\
            289118 & S0/eS & S0/eS & S0/eS & - \\
            289168 & E/S0 & E/S0 & E & - \\
            289170 & S0 & S0 & S0 & - \\
            290328 & eS & eS & eS & - \\
            296639 & eS/lS & eS/lS & lS & - \\
            296677 & S0/eS & S0/eS & eS & - \\
            296742 & lS & lS & ? & NR \\
            296829 & S0/eS & S0/eS & eS & - \\
            297053 & S0/eS & S0/eS & S0/eS & - \\
            298572 & eS/lS & eS/lS & eS/lS & - \\
            298590 & E & E & E & - \\
            298591 & S0/eS & S0/eS & eS & - \\
            298862 & eS/lS & eS/lS & eS/lS & - \\
            298980 & ? & ? & E & - \\
            300406 & E/S0 & E/S0 & E & - \\
            300411 & E & E & E & - \\
            300691 & E & E & E & - \\
            300787 & S0 & S0 & S0 & - \\
            301191 & S0/eS & S0/eS & eS & - \\
            301199 & S0/eS & S0/eS & S0/eS & - \\
            301346 & eS/lS & eS/lS & lS & - \\
            301799 & ? & ? & ? & - \\
            301937 & E/S0 & Merger & E & Merger \\
            302684 & S0 & S0 & S0 & - \\
            303099 & eS & eS & eS & - \\
            319060 & S0 & S0 & S0 & - \\
            319067 & S0 & S0 & S0 & - \\
            319070 & E/S0 & E/S0 & E/S0 & - \\
            319157 & S0/eS & S0/eS & S0 & - \\
            319196 & E/S0 & E/S0 & ? & - \\
            319197 & S0/eS & S0/eS & S0 & - \\
            319393 & lS & lS & lS & - \\
            319397 & eS & eS & eS & - \\
            319400 & eS & eS & eS & - \\
            319402 & S0 & S0 & E & - \\
            319452 & S0 & S0 & S0 & - \\
            320281 & lS & lS & lS & - \\
            321035 & S0 & S0 & S0 & - \\
            321146 & E/S0 & E/S0 & E & - \\
            323242 & ? & ? & ? & - \\
            323329 & eS & eS & eS & - \\
            323558 & S0/eS & S0/eS & S0/eS &  \\
            323566 & E & E & E & - \\
            323854 & S0/eS & S0/eS & S0/eS & Merger \\
            324323 & lS & lS & lS & - \\
            324351 & S0/eS & S0/eS & S0/eS & - \\
            325007 & eS & eS & eS & - \\
            325321 & S0 & S0 & S0 & - \\
            345820 & S0/eS & S0/eS & S0 & - \\
            346046 & S0 & S0 & S0 & - \\
            346623 & E/S0 & E/S0 & S0 & - \\
            346837 & E & E & E & - \\
            346839 & S0 & S0 & S0 & - \\
            346887 & E/S0 & E/S0 & E & Merger \\
            346890 & S0 & S0 & S0 & - \\
            346892 & eS & eS & eS & - \\
            346894 & S0/eS & S0/eS & eS & - \\
            347432 & S0 & S0 & Embedded Disc & - \\
            347470 & eS & eS & eS & - \\
            365334 & eS & eS & eS & - \\
            371145 & S0/eS & S0/eS & S0/eS & - \\
            371171 & E & E & E & - \\
            371172 & E & E & E & - \\
            371789 & eS & eS & eS & - \\
            371889 & S0/eS & S0/eS & eS & - \\
            371976 & S0 & S0 & S0 & - \\
            372105 & S0/eS & S0/eS & S0 & - \\
            372133 & E/S0 & E/S0 & S0 & - \\
            372408 & S0 & S0 & S0 & - \\
            372446 & ? & ? & E & Merger \\
            372452 & E/S0 & E/S0 & E/S0 & - \\
            372568 & eS/lS & eS/lS & eS & - \\
            372599 & eS/lS & eS/lS & eS & - \\
            373105 & S0/eS & S0/eS & eS & - \\
            375564 & S0 & S0 & S0 & - \\
            375615 & E/S0 & E/S0 & E & - \\
            376001 & ? & ? & ? & - \\
            376703 & E & E & E & - \\
            377666 & S0/eS & S0/eS & eS & - \\
            378104 & S0 & S0 & S0 & - \\
            380697 & S0 & S0 & S0 & - \\
            381156 & S0/eS & S0/eS & S0 & - \\
            381157 & S0 & S0 & S0 & - \\
            381160 & S0 & S0 & S0 & - \\
            381219 & E/S0 & E/S0 & eS & - \\
            382158 & S0 & S0 & S0 & - \\
            382505 & E & E & E & - \\
            382563 & E/S0 & E/S0 & E & - \\
            382631 & eS/lS & eS/lS & lS & - \\
            382659 & S0 & S0 & S0 & - \\
            383259 & eS & eSB & eSB & - \\
            386268 & S0 & S0 & S0 & - \\
            386316 & S0/eS & S0/eS & S0 & - \\
            386366 & eS & eS & eS & - \\
            386673 & E/S0 & E/S0 & E/S0 & - \\
            386722 & E/S0 & E/S0 & S0 & - \\
            387042 & S0 & S0 & S0 & - \\
            387080 & S0/eS & S0/eS & S0/eS & - \\
            387675 & S0/eS & S0/eS & S0 & - \\
            387715 & S0 & S0 & S0 & - \\
            388476 & eS & eS & eS & - \\
            388513 & E/S0 & E/S0 & E & - \\
            388552 & S0 & S0 & S0 & - \\
            396584 & eS & eS & eS & - \\
            396607 & eS & eS & eS & - \\
            396615 & S0/eS & S0/eS & eS & - \\
            396709 & eS/lS & eS/lS & eS/lS & - \\
            396781 & S0/eS & S0/eS & eS & - \\
            396833 & eS & eS & eS & - \\
            402988 & ? & ? & E/S0 & - \\
            417392 & lS & lS & lS & - \\
            417433 & eS & eS & eS & - \\
            417678 & eS/lS & eS/lS & lS & - \\
            418630 & S0 & S0 & S0 & - \\
            418693 & S0/eS & S0/eS & eS & - \\
            418725 & eS & eS & eS & - \\
            419294 & S0 & S0 & S0 & - \\
            419335 & S0 & S0 & S0 & - \\
            419345 & S0 & S0 & S0 & - \\
            422286 & eS & eSB & eSB & - \\
            422289 & eS/lS & eS/lS & ? & NR \\
            422292 & S0 & S0 & S0 & - \\
            422359 & S0/eS & S0/eS & S0 & - \\
            422389 & E & E & E & - \\
            422406 & E & E & E & - \\
            422443 & S0/eS & S0/eS & S0 & - \\
            422446 & S0/eS & S0/eS & eS & - \\
            422610 & lS & lS & lS & - \\
            457615 & S0/eS & S0/eS & S0 & - \\
            460380 & eS & eS & eS & - \\
            460756 & eS & eS & eS & - \\
            460767 & S0/eS & S0/eS & S0 & - \\
            463148 & E/S0 & E/S0 & E & - \\
            463434 & eS/lS & eS/lS & eS & - \\
            463641 & lS & lS & lS & - \\
            485047 & S0/eS & S0/eS & S0 & - \\
            485127 & eS & eS & eS & - \\
            485688 & eS/lS & eS/lS & eS & - \\
            485834 & eS/lS & eS/lS & eS/lS & - \\
            485882 & eS & eS & eS & - \\
            485883 & E/S0 & E/S0 & E & - \\
            485921 & E & E & E & - \\
            485924 & eS & eS & eS & - \\
            486276 & eS & eS & eS & - \\
            491508 & S0 & S0 & ? & - \\
            491887 & eS & eS & eS & - \\
            491960 & eS & eS & eS & - \\
            492384 & eS & eS & eS & - \\
            492883 & S0/eS & S0/eS & S0/eS & - \\
            492941 & E & E & E & - \\
            493838 & S0 & S0 & S0 & - \\
            497035 & E/S0 & E/S0 & S0 & Interacting \\
            504730 & S0 & S0 & S0 & - \\
            504877 & S0 & S0 & S0 & - \\
            504879 & E/S0 & E/S0 & E & - \\
            505316 & eS & eS & eS & - \\
            505341 & S0 & S0 & Embedded Disc & - \\
            505765 & E & E & E & - \\
            505828 & eS & eS & lS & - \\
            506119 & E & E & E & - \\
            508138 & S0/eS & S0/eS & S0 & - \\
            508144 & S0/eS & S0/eS & S0 & - \\
            508184 & S0/eS & S0/eS & eS & - \\
            508312 & S0/eS & S0/eS & S0 & - \\
            508315 & S0/eS & S0/eS & S0/eS & - \\
            508568 & S0/eS & S0/eS & eS & - \\
            508593 & E/S0 & E/S0 & E & - \\
            508751 & E/S0 & E/S0 & E & - \\
            509075 & E & E & E & - \\
            511456 & eS/lS & eS/lS & eS & - \\
            511864 & ? & ? & E/S0 & - \\
            511890 & S0/eS & S0/eS & eS & - \\
            511892 & S0 & S0 & S0 & - \\
            511921 & lS & lS & lS & - \\
            514022 & S0/eS & S0/eS & lS & - \\
            514212 & lS & lS & lS & - \\
            517244 & S0/eS & S0/eS & S0 & - \\
            517273 & lS & lS & lS & - \\
            517302 & S0/eS & eSB & eSB & - \\
            518829 & E/S0 & E/S0 & S0 & - \\
            519141 & E/S0 & E/S0 & S0 & - \\
            521736 & eS/lS & eS/lS & lS & - \\
            527984 & eS & eS & eS & - \\
            534654 & eS/lS & eS/lS & E & NR \\
            534710 & eS/lS & eS/lS & ? & - \\
            534763 & S0 & S0 & ? & NR \\
            534904 & S0 & S0 & S0 & - \\
            535387 & eS & eS & eS & - \\
            535417 & lS & lS & lS & - \\
            536454 & E & E & E & - \\
            536463 & S0 & S0 & S0 & - \\
            536963 & E/S0 & E/S0 & S0 & - \\
            537467 & S0 & S0 & ? & Merger \\
            543501 & S0 & S0 & S0 & - \\
            543552 & S0/eS & S0/eS & S0 & - \\
            543895 & eS/lS & eS/lS & eS/lS & KDC? \\
            545903 & S0 & S0 & Embedded Disc & - \\
            548835 & E/S0 & E/S0 & E & NR \\
            548872 & S0/eS & S0/eS & S0 & - \\
            548946 & eS & eSB & eSB & - \\
            549070 & E/S0 & E/S0 & E & - \\
            549313 & ? & ? & lS & - \\
            550328 & eS & eS & eS & - \\
            550780 & eS & eS & eS & - \\
            550781 & S0 & S0 & S0 & - \\
            551332 & eS/lS & eS/lS & lS & - \\
            559490 & E & E & E & - \\
            560407 & eS/lS & eS/lS & ? & - \\
            560982 & E/S0 & E/S0 & E & - \\
            561488 & S0/eS & S0/eS & S0 & - \\
            561856 & eS & eS & eS & - \\
            569466 & eS & eS & eS & - \\
            569555 & S0/eS & S0/eS & lS & - \\
            570203 & E/S0 & E/S0 & E & - \\
            570206 & S0/eS & S0/eS & eS & - \\
            570207 & S0 & S0 & S0 & - \\
            570227 & eS & eS & eS & - \\
            570240 & S0 & S0 & S0 & - \\
            573617 & S0/eS & S0/eS & S0 & - \\
            573621 & eS & eS & eS & - \\
            574285 & eS & eS & eS & - \\
            574617 & eS & eS & lS & - \\
            575339 & E & E & E & - \\
            575532 & eS & eS & eS & - \\
            584013 & eS & eSB & eSB & - \\
            585196 & E/S0 & E/S0 & E/S0 & - \\
            585215 & E & E & E & - \\
            586330 & S0 & S0 & S0 & - \\
            592294 & S0 & S0 & S0 & - \\
            592858 & S0/eS & S0/eS & eS & - \\
            593117 & eS/lS & eS/lS & eS & - \\
            593642 & S0/eS & S0/eS & ? & - \\
            593813 & S0 & S0 & S0 & - \\
            594053 & S0/eS & S0/eS & S0 & - \\
            594834 & eS & eS & eS & - \\
            594906 & lS & lS & lS & - \\
            594990 & S0/eS & S0/eS & eS & - \\
            595060 & eS & eS & lS & - \\
            595088 & lS & lS & lS & - \\
            598999 & eS & eS & eS & - \\
            599846 & eS/lS & eS/lS & eS & - \\
            600743 & E/S0 & E/S0 & S0 & - \\
            600882 & E/S0 & E/S0 & S0 & - \\
            600915 & S0 & S0 & S0 & - \\
            600973 & E/S0 & E/S0 & E & NR \\
            601102 & S0/eS & S0/eS & S0 & - \\
            609396 & eS/lS & eS/lS & lS & - \\
            609948 & S0/eS & eS & eS & - \\
            610054 & eS/lS & eS/lS & eS & - \\
            610396 & S0/eS & S0/eS & eS & - \\
            610404 & S0 & S0 & S0 & - \\
            610474 & eS & eSB & eSB & - \\
            617989 & S0/eS & S0/eS & eS & - \\
            618108 & eS & eS & eS & - \\
            618142 & S0 & S0 & E & - \\
            618143 & S0/eS & S0/eS & S0 & - \\
            618144 & S0 & S0 & S0 & - \\
            618151 & E & E & E & - \\
            618904 & S0/eS & S0/eS & S0 & - \\
            618906 & S0/eS & S0/eS & eS & - \\
            618912 & E/S0 & E/S0 & E & - \\
            618930 & E/S0 & E/S0 & E & - \\
            618935 & lS & lS & lS & - \\
            618937 & E/S0 & E/S0 & S0 & - \\
            618938 & S0 & S0 & S0 & - \\
            618941 & S0 & S0 & S0 & - \\
            618942 & E & E & E & - \\
            618947 & E/S0 & E/S0 & E/S0 & - \\
            618952 & S0/eS & S0/eS & eS & - \\
            618976 & S0 & S0 & eS & - \\
            618992 & eS & eS & ? & Merger, Interacting \\
            619095 & eS/lS & eS/lS & eS & - \\
            619097 & eS & eS & eS & - \\
            622596 & eS & eS & eS & - \\
            622694 & eS & eS & eS & - \\
            622770 & S0/eS & S0/eS & S0/eS & - \\
            622787 & S0 & S0 & S0 & - \\
            623043 & S0 & S0 & S0 & - \\
            623143 & S0/eS & S0/eS & S0/eS & - \\
            623161 & eS & eS & eS & - \\
            623272 & eS & eS & eS & - \\
            663440 & E/S0 & E/S0 & S0 & - \\
            3628218 & E/S0 & E/S0 & S0 & - \\
            3630097 & eS & eS & eS & - \\
            3630943 & S0/eS & S0/eS & eS & - \\
            3631885 & S0/eS & S0/eS & S0/eS & - \\
            3894635 & eS & eS & eS & - \\
            3895503 & ? & ? & E & - \\
            3895823 & eS/lS & eS/lS & eS & - \\
            3896912 & lS & lS & lS & - \\
            3897281 & S0/eS & S0/eS & S0/eS & - \\
            \hline
    \end{longtable}